\documentclass[envcountsect]{svmult}

\usepackage[utf8]{inputenc}      
\usepackage[T1]{fontenc}         
\usepackage[leqno]{amsmath}
\usepackage{amssymb}

\usepackage{eucal}
\usepackage{textcomp}
\usepackage{paralist}
\usepackage{appendix}
\usepackage[dvipsnames]{xcolor}  
\usepackage{tikz}
\usetikzlibrary{arrows,calc,intersections,positioning,shapes.geometric}

\usepackage{amsthm}  \swapnumbers

\numberwithin{equation}{section}
\makeatletter
\def\ds@numart{\@numarttrue
  \@takefromreset{figure}{chapter}%
  \@takefromreset{table}{chapter}%
  \@takefromreset{equation}{chapter}%
  \def\thesection{\@arabic\c@section}%
  \def\thefigure{\@arabic\c@figure}%
  \def\thetable{\@arabic\c@table}%
  \def\theequation{\thesection.\arabic{equation}}%
  \def\thesubequation{\arabic{equation}\alph{subequation}}}
\makeatother

\makeatletter
  \def\swappedhead@plain#1#2#3{%
    \thmnumber{(\textup{#2})}
    \thmname{\@ifnotempty{#2}{~}\textup{#1}}
    \thmnote{ {\textup{(#3)}}}}
  \let\swappedhead\swappedhead@plain
\makeatother

\newtheorem{mytheo}[equation]{Theorem}
\newtheorem{myprop}[equation]{Proposition}
\newtheorem{mycoro}[equation]{Corollary}

\newtheorem{mytheodefi}[equation]{Theorem, Definition}
\newtheorem{mypropdefi}[equation]{Proposition, Definition}
\theoremstyle{definition}
\newtheorem{mydefi}[equation]{Definition}
\newtheorem{mydefis}[equation]{Definitions}
\newtheorem{myexam}[equation]{Example}
\newtheorem{myrema}[equation]{Remark}
\newtheorem{myremas}[equation]{Remarks}

\usepackage{Z-tikz}

\usepackage{txfonts}
 
\usepackage{bm}     

\usepackage[backend=biber,bibstyle=Z-bib,citestyle=alphabetic]{biblatex}

\usepackage[colorlinks=true,%
            linkcolor=RoyalBlue,%
            citecolor=RedViolet,%
            urlcolor=RoyalBlue]{hyperref}   

\let\polishl\l

  \newcommand\CC{{\mathbf{C}}}         
  \newcommand\RR{{\mathbf{R}}}         
  \newcommand\ZZ{{\mathbf{Z}}}         

  \newcommand\Lie{\mathfrak}
  \newcommand\LA{{\Lie{a}}}            
  \newcommand\LB{{\Lie{b}}}            
  \newcommand\LC{{\Lie{c}}}            
  \newcommand\LG{{\Lie{g}}}            
  \newcommand\LH{{\Lie{h}}}            
  \newcommand\LK{{\Lie{k}}}            
  \newcommand\LT{{\Lie{t}}}            

  \newcommand\inv{^{-1}}                
  \DeclareMathOperator\Ad{Ad}          
  
  \let\Re\relax
  \DeclareMathOperator\Re{Re}           

  \newcommand\+{^+}
  \newcommand{\<}{\langle}
\renewcommand{\>}{\rangle}
\renewcommand\b{\mathrm b}             
  \newcommand\ch{^{\hspace{.5pt}\scriptscriptstyle{\bm{\vee}}}} 
  \newcommand{\compsign}[1]{\vcenter{\hbox{$#1\circ$}}} 
  \newcommand{\comp}{\mathbin{\mathchoice 
   {\compsign\scriptstyle}
   {\compsign\scriptstyle} 
   {\compsign\scriptscriptstyle}
   {\compsign\scriptscriptstyle}}}
\renewcommand\comp\circ
\renewcommand\d{{\delta}}
  \newcommand\GNS[1]{\operatorname{GNS}_{#1}}
\renewcommand\j{{\mathrm{j}}}
  \newcommand\sub[1]{_{\smash{\raisebox{.5pt}{$\scriptstyle #1$}}}}
  \newcommand\suba{_{|\LA}}
\DeclareMathOperator\Aff{Aff}
\DeclareMathOperator\aut{aut}
\DeclareMathOperator\Aut{Aut}
\DeclareMathOperator\Conv{Conv}
\let\div\relax
\DeclareMathOperator\div{div}
\DeclareMathOperator\Hom{Hom}
\DeclareMathOperator\ind{ind}
\DeclareMathOperator\Ind{Ind}
\DeclareMathOperator\rot{\mathbf{curl}}

\showhyphens{Heisenberg hamiltonian Helmholtz}

\showhyphens{decomposition diffeomorphism dimensional distributions equivariance imprimitivity isomorphism Kazhdan modules momentum periodic probability representation represents represented submersion submersions theoretic}

\bibliography{biblio}

\usepackage{helvet}         
\usepackage{makeidx}         
\usepackage{graphicx}        
\usepackage{multicol}        
\usepackage[bottom]{footmisc}

\makeindex             

\begin{document}

\motto{À la mémoire de Jean-Marie Souriau}
\title*{Localized Quantum States}
\author{Fran\c cois Ziegler}
\institute{Fran\c cois Ziegler \at Department of Mathematical Sciences, Georgia Southern University, Statesboro, GA 30460-8093, USA, \email{fziegler@georgiasouthern.edu}}

\maketitle

\abstract{Let $X$ be a symplectic manifold and $\Aut(L)$ the automorphism group of a Kostant-Souriau line bundle on $X$. \emph{Quantum states for} $X$, as defined by J.\nobreakdash-M.~Souriau in the 1990s, are certain positive-definite functions on $\Aut(L)$ or, less ambitiously, on any ``large enough'' subgroup $G\subset\Aut(L)$. This definition has two major drawbacks: when $G=\Aut(L)$ there are no known examples; and when $G$ is a Lie subgroup the notion is, as we shall see, far from selective enough. In this paper we introduce the concept of a quantum state \emph{localized at} $Y$, where $Y$ is a coadjoint orbit of a subgroup $H$ of $G$. We show that such states exist, and tend to be unique when $Y$ has lagrangian preimage in $X$. This solves, in a number of cases, A.~Weinstein's ``fundamental quantization problem'' of attaching state vectors to lagrangian submanifolds.}

\vspace{2ex}

\setcounter{minitocdepth}{1}
\dominitoc

\section{Introduction: The quantization problem}

\emph{Quantum mechanics is a unitary representation of the symmetry group of classical mechanics---or a large subgroup thereof}. This prescription, which infinitesimally goes back to Dirac \cite[§21]{Dirac:1930}, first became precise in 1965 when Kostant and Souriau constructed the symmetry group in question: namely, it is the automorphism group of a Kostant-Souriau line (or circle) bundle, $L$, over the symplectic manifold $X$ which models the classical mechanical system under consideration.


\begin{myexam}[the plane] 
   \label{R2}
   Let $X$ be $\RR^2$ with points $x=(p,q)$ and 2-form $\omega=dp\wedge dq$. Then $L$ is $X\times\CC$ with points $\xi=(x,z)$, projection $\xi\mapsto x$, connection 1-form $\varpi=pdq + dz/\I z$, and hermitian structure $|\xi|=|z|$. An automorphism, $g\in\Aut(L)$, is a diffeomorphism of the form
   \begin{equation}
      \label{Aut(L)}
      g(x,z) = \bigl(s(x), z\E^{\I S(x)}\bigr)
   \end{equation}
   where $s$ is a symplectomorphism of $X$ and the function $S$ is determined up to an additive constant by the condition that $pdq-s^*(pdq)=dS$. The Lie algebra $\aut(L)$ of infinitesimal automorphisms of $L$ is isomorphic to the Poisson bracket algebra $\mathrm{C}^\infty(X)$: to any $(\varpi, |\cdot|)$-preserving vector field $Z$ we can attach the function $H(x)=\varpi(Z(\xi))$ called its \emph{\textbf{hamiltonian}}, and conversely any $H\in\mathrm{C}^\infty(X)$ gives rise to the infinitesimal automorphism 
   \begin{equation}
      \label{aut(L)}
      Z(x,z) = \bigl(\eta(x),\I z\ell(x))
   \end{equation}
   where $\eta = (-\partial H/\partial q, \partial H/\partial p)$ is the symplectic gradient of $H$, and $\ell = H-p\partial H/\partial p$. (This isomorphism is established in greater generality in \cites{Kostant:1970}{Souriau:1970}; in the case at hand it was already known to Lie and Van Hove \cites[p.\,270]{Lie:1890}[§5]{Van-Hove:1951}.)
\end{myexam}

Given a symplectic manifold $X$ and a Kostant-Souriau line bundle $L$ over it, one would now of course like to know \emph{which} representation(s) of $\Aut(L)$---or of subgroups thereof---furnish the quantum theory. As $\Aut(L)$-invariant ``polarizations'' are not available, Souriau was led to propose instead the following axiomatic, polarization-independent definition.

\begin{mydefi}[\cites{Souriau:1988}{Souriau:1990a}{Souriau:1992}]
   \label{QR1}
   A \emph{\textbf{quantum representation}} (of $\Aut(L)$, for $X$) is a unitary $\Aut(L)$-module $\mathcal H$ such that, for every unit vector $\varphi\in\mathcal H$, the matrix coefficient $m(g)=(\varphi,g\varphi)$ satisfies
   \begin{equation}
      \label{inequalities1}
      \Bigl|\sum_{j=1}^n c_jm(\exp(Z_j))\Bigr|
      \leqslant
      \sup_{x\in X}\Bigl|\sum_{j=1}^n c_j\E^{\I H_j(x)}\Bigr|
   \end{equation}
   for all choices of an integer $n$, complex numbers $c_1,\dots,c_n$ and complete, commuting vector fields $Z_1,\dots,Z_n\in\aut(L)$ with respective hamiltonians $H_1,\dots,H_n$. (Here $\exp(Z_j)\in\Aut(L)$ denotes the time 1 flow of the complete vector field $Z_j\in\aut(L)$.)
   As we shall see in §2, \eqref{inequalities1} can be reformulated (after \cite{Ziegler:1996b}) as requiring that
\begin{equation}
   \label{geomcrit0}
   \begin{gathered}
	   \text{\emph{the quantum spectrum of `commuting observables'}}\\[-.7ex]
	   \text{\emph{is concentrated on their classical range, suitably compactified.}}
   \end{gathered}
\end{equation}
\end{mydefi}
\noindent
The \emph{\textbf{problem of geometric quantization}}, in the words of \cite[p.\,74]{Souriau:1984a}, is now to find a quantum representation of $\Aut(L)$; or equivalently---see \eqref{GNS_quantum}---to find a \emph{state} $m$ of $\Aut(L)$ satisfying \eqref{inequalities1}. This is a tall order, which we will not address here beyond observing that 1º) the ``obstruction theorem'' of \cite{Van-Hove:1951} does \emph{not} prove its impossibility, yet 2º) the solution is not the so-called prequantization representation (also introduced in \cite{Van-Hove:1951}; see §2). Instead we shall study, as the start of this introduction suggests, states and representations of \emph{Lie subgroups} $G\subset\Aut(L)$ that satisfy the inequalities induced by \eqref{inequalities1}. The main points of our investigation are as follows:
\begin{enumerate}
   \item[--] In §3 we show that Souriau's resulting notions of quantum state and representation (of a Lie group $G$, for one of its coadjoint orbits $X$) are by themselves not selective enough, because the compactification in \eqref{geomcrit0} can fail utterly to distinguish between coadjoint orbits.
   \item[--] In \cite{Ziegler:1996b} this was remedied by \emph{suppressing} this compactification. Here in contrast we take it seriously, because we find that it (and only it) makes room for interesting, \emph{localized} states---defined in §4 by the property that their further restriction to a Lie subgroup $H\subset G$ is quantum for a coadjoint orbit $Y$ of $H$.
   \item[--] In §5 we prove existence and uniqueness, whenever $G$ is a nilpotent Lie group and $\LH$ is what Kirillov called a maximal subordinate subalgebra to $x\in\LG^*$, of a quantum state for $X=G(x)$ localized at $Y=\{x_{|\LH}\}$. This vastly generalizes states of the Heisenberg group discussed in \cite{Beaume:1974,Ashtekar:2003}.
   \item[--] In §6 we prove existence and uniqueness, whenever $G$ is a compact Lie group, $T$ a maximal torus and $x$ an integral, $T$-fixed point in $\LG^*$, of a quantum state for $X=G(x)$ localized at $Y=\{x_{|\LT}\}$. The resulting Gel'fand-Na{\u\i}mark-Segal representation is the irreducible one with highest weight $\lambda=x_{|\LT}$.
   \item[--] In §7 we prove existence and sometimes uniqueness of several quantum states of Euclid's group for the coadjoint orbit $X$ relevant in geometrical optics, localized at orbits $Y$ having lagrangian preimages in $X$. These states provide legitimate hilbertian models of the physicists' \emph{plane}, \emph{spherical} and \emph{cylindrical} waves.
\end{enumerate}
Finally the Appendix collects a number of known facts on positive-definite functions, states, and unitary representations of groups used throughout the paper.

\section{Prequantization is not quantum}

We start by giving the promised geometric recasting \eqref{geomcrit0} of inequalities \eqref{inequalities1}. To this end, let us agree to call \emph{\textbf{perspective on}} $X$ any finite-dimensional, commutative subalgebra $\LA$ of $\aut(L)$ consisting of complete vector fields. Given such an $\LA$ and $x\in X$, write $x\suba$ for the character $Z\mapsto \E^{\I H(x)}$ of $\LA$, where $H$ is the hamiltonian of $Z$; and regard $x\mapsto x\suba$ as a map of $X$ to the (compact) Pontryagin dual $\hat\LA$ of the \emph{discretized} additive group $\LA$. Then we have:

\begin{mytheo}
   \label{geomcrit1}
   A unitary $\Aut(L)$-module $\mathcal H$ is a quantum representation for $X$ if and only if for each unit vector $\varphi\in\mathcal H$ and each perspective $\LA$ on $X$\textup, the state $(\varphi,\exp\suba(\,\cdot\,)\varphi)$ of $\LA$ has its spectral measure concentrated on the closure of $X\suba$ in $\hat\LA$.
\end{mytheo}

(We refer to the Appendix for the notions of \emph{state} \eqref{state} and \emph{spectral measure} \eqref{Bochner}. The closure of $X\suba$ in $\hat\LA$ is the compactification mentioned in \eqref{geomcrit0}, and can be viewed as an abstract device allowing us to treat the inequalities \eqref{inequalities1} all at once; the group $\hat\LA$ itself is known as the \emph{Bohr compactification} $\b\LA^*$ of the ordinary dual $\LA^*$ of $\LA$: see \cite[{}26.11]{Hewitt:1963}.)

\begin{proof}
   Suppose that $\mathcal H$ satisfies \eqref{inequalities1}, and let $\LA$ be a perspective on $X$. Then the function $(\varphi,\exp\suba(\,\cdot\,)\varphi)=m\comp\exp\suba$ is the pull-back of a state by a group homomorphism, hence is a state as one readily verifies.
   By Bochner's theorem \eqref{Bochner} this state has a spectral measure $\nu$ so that $(m\comp\exp\suba)(Z)=\int_{\hat\LA}\chi(Z)\,d\nu(\chi)$. Now \eqref{inequalities1} says that we have $|\nu(f)|\leqslant\sup_{x\in X}|f(x\suba)|$, or in other words
   \begin{equation}
      \label{meanvalue}
   	|\nu(f)|\leqslant\sup_{\chi\in X\suba}|f(\chi)|,
   \end{equation}
   for every trigonometric polynomial $f(\chi)=\sum_j c_j\chi(Z_j)$ with $c_j\in\CC$, $Z_j\in\LA$. By Stone-Weierstrass, these are uniformly dense in the continuous functions on $\hat\LA$, so therefore \eqref{meanvalue} still holds for all continuous $f$. In particular if $f$ vanishes on the closure $\b X\suba$ of $X\suba$ in $\hat\LA$ then $\nu(f)=0$, which is to say that
   \begin{equation}
      \label{support}
   	\operatorname{supp}(\nu)\subset\b X\suba,
   \end{equation}
   or in other words, that $\nu$ is concentrated on $\b X\suba$ \cite[n$^{\textrm o}$ V.5.7]{Bourbaki:1967b}.
   
   Conversely let $c_j$ and $Z_j$ be given as in Definition \eqref{QR1}. Then the $Z_j$ span a perspective $\LA$ on $X$, and $f(\chi) =\sum_j c_j\chi(Z_j)$ defines a continuous function on $\hat\LA$. Assuming \eqref{support} for $\LA$, the mean value inequality gives us \eqref{meanvalue} and hence \eqref{inequalities1}.
\end{proof}

\begin{myexam}[continued]
   \label{R2bis}
   The space of $\mathrm L^2$ sections of the line bundle $L$ of \eqref{R2} is naturally a unitary $\Aut(L)$-module, often called the \emph{prequantization representation}. Identifying sections $\sigma$ with functions $\varphi\in\mathrm L^2(X)$ by writing $\sigma(x)=(x,\varphi(x))$, the action of an automorphism \eqref{Aut(L)} reads
   \begin{equation}
      \label{prequantization}
      (g\varphi)(x) = \E^{\I S(s\inv(x))}\varphi(s\inv(x)).
   \end{equation}
   We claim:
\end{myexam}

\begin{myprop}
   \label{propR2}
   The prequantization representation \eqref{prequantization} of $\Aut(L)$ in $\mathrm L^2(X)$ is not quantum for $X$.
\end{myprop}

\begin{proof}
   We consider the hamiltonian $H(p,q)=\sin p$. It gives rise to an infinitesimal automorphism \eqref{aut(L)} whose flow writes $\E^{tZ}(p,q,z) = \bigl(p, q+t\cos p,z\E^{\I t(\sin p - p\cos p)}\bigr)$. The resulting action \eqref{prequantization} on sections is 
   \begin{equation}
      \label{flowsin}
   	(\E^{tZ}\varphi)(p,q) =\E^{\I t(\sin p - p\cos p)}\varphi(p, q-t\cos p).
   \end{equation}
   In order to compute its spectral measure, we introduce the partial Fourier transform $\hat\varphi(p,k) = {\scriptstyle\surd(2\pi)\inv}\!\! \int \E^{\I kq}\varphi(p,q)\,dq$ on which the transported action becomes
   \begin{equation}
      (\E^{tZ}\hat\varphi)(p,k) = \E^{\I t(\sin p + (k-p)\cos p)}\hat\varphi(p, k).
   \end{equation}
   This demonstrates that the spectral measure \eqref{Bochner} of $tZ\mapsto (\varphi,\E^{tZ}\varphi)$ is the image of $|\hat\varphi(p,k)|^2dp\,dk$ by the map $(p,k)\mapsto \sin p + (k-p)\cos p$. Now if \eqref{prequantization} were quantum for $X$, then by Theorem \eqref{geomcrit1} this image measure would be always concentrated on the range $[-1,1]$ of $H$ (in $\RR$, which we have identified with the dual $\LA^*\subset\hat\LA$ of the perspective $\LA=\RR Z$); but this is clearly not the case.
\end{proof}

\begin{myremas}
   It is comforting to see Definition \eqref{QR1} eliminate the prequantization representation \eqref{prequantization}, which physicists since Van Hove \cite{Van-Hove:1951} have rejected as ``too big''. But let us emphasize that it does so for \emph{different reasons}.
   
   For Van Hove, the trouble with \eqref{prequantization} is that restricting it to the automorphisms $g(p,q,z)=\bigl(p+b, q+c, z\E^{-\I(a + bq)}\bigr)$ by which the Heisenberg group
   \begin{equation}
      \label{Heisenberg}
      G=\left\{g=
      \begin{pmatrix}
         \,1 & \,b\, & a\,\\
           & 1 & c\,\\
           &   & 1\,
      \end{pmatrix}
      : a,b,c\in\RR
      \right\}
   \end{equation}
   acts on $L$, produces a representation 
   \begin{equation}
      \label{reducible}
      (g\varphi)(p,q) = \E^{-\I a}\E^{-\I b(q-c)}\varphi(p-b,q-c)
   \end{equation}
   of $G$ which is \emph{reducible} and thus not equivalent to the Schrödinger representation. (Van Hove went on to demand that any acceptable representation of $\Aut(L)$ be irreducible on $G$, and then to prove his famous ``obstruction theorem'' that no such representation could possibly exist.)

   Definition \eqref{QR1}, in contrast, imposes no such irreducibility condition (we fully expect that a representation satisfying it will \emph{not} be irreducible on $G$); and the sense in which it declares \eqref{prequantization} ``too big'' is purely \emph{spectral}: this representation assigns too large a spectrum to the bounded quantity $\sin p$. Another advantage is that Definition \eqref{QR1} excludes \emph{more} undesired representations---such as the following, once proposed by Gotay and rejected by Velhinho (see \cites{Velhinho:1998,Gotay:2000}).
\end{myremas}

\begin{myexam}[the 2-torus]
   Consider the pair $L\to X$ of \eqref{R2} and three numbers $A, B, C$ with $A = BC = 2\pi$. Then a particular Kostant-Souriau line bundle over the torus $\dot X = \RR^2/(B\ZZ\times C\ZZ)$ is the quotient $\dot L = L/\Gamma$ of $L$ by the action of the subgroup $(a,b,c)\in A\ZZ\times B\ZZ\times C\ZZ$ of \eqref{Heisenberg}. Its $\mathrm L^2$ sections can be identified with functions on $X$ that satisfy
   \begin{equation}
      \label{Bloch}
      \varphi(p+b,q+c) = \E^{-\I bq}\varphi(p,q)
   \end{equation}
   for all $(b,c)\in B\ZZ\times C\ZZ$, and are square integrable over any rectangle of size $B\times C$. Specializing to $C=1$, the flow with hamiltonian $\sin p$ on $L$ commutes with $\Gamma$ and so descends to act on $\dot L$ and on its sections \eqref{Bloch} by the same formula \eqref{flowsin} as before. Arguing much as in \eqref{propR2} (with a Fourier series replacing the Fourier transform), one readily obtains:
\end{myexam}

\begin{myprop}
   The prequantization representation of $\Aut(\dot L)$ in $\mathrm L^2$ sections of $\dot L\to\dot X$ is not quantum for the 2-torus $\dot X$.
\end{myprop}

\section{Quantum states for coadjoint orbits}

It is unknown whether any representation satisfying Definition \eqref{QR1} exists beyond the simple case where $X$ is a single point. So, heeding the advice at the start §1, we shall look instead for representations of \emph{Lie subgroups} of $\Aut(L)$, where $L\to X$ is a Kostant-Souriau line bundle; or equivalently (see \eqref{GNSS}), for \emph{states} of Lie groups $G$ having a smooth action $G\to\Aut(L)$.

Such an action has a canonical moment map $\Phi:X\to\LG^*$, where $\<\Phi(\,\cdot\,),Z\>$ is the hamiltonian of the image of $Z\in\LG$ in $\aut(L)$. We will regard $G$ as ``large enough'' if these hamiltonians separate points of $X$; then the moment map is one-to-one, and we may as well assume that $X$ is a coadjoint orbit of $G$. Thus we come to:

\begin{mydefi}[\cites{Souriau:1988}{Souriau:1990a}{Souriau:1992}]
   \label{QR2}
   Let $X$ be a coadjoint orbit of the Lie group $G$. A \emph{\textbf{quantum state}} (of $G$, for $X$) is a state $m$ of $G$ such that
   \begin{equation}
      \label{inequalities2}
      \Bigl|\sum_{j=1}^n c_jm(\exp(Z_j))\Bigr|
      \leqslant
      \sup_{x\in X}\Bigl|\sum_{j=1}^n c_j\E^{\I\<x,Z_j\>}\Bigr|
   \end{equation}
   for all choices of an integer $n$, complex numbers $c_j$, and \emph{commuting} $Z_j$ in the Lie algebra $\LG$ of $G$. A \emph{\textbf{quantum representation}} (of $G$, for $X$) is a unitary $G$\nobreakdash-module $\mathcal H$ such that, for every unit $\varphi\in\mathcal H$, the function $m(g)=(\varphi,g\varphi)$ is a quantum state.
\end{mydefi}

\begin{mytheo}[{\cite[{}5.2b]{Souriau:1988}}]
   \label{GNS_quantum}
   A state $m$ of $G$ is quantum for $X$ if and only if the resulting Gel'fand-Na{\u\i}mark-Segal representation\textup, $\GNS{m}$ \eqref{GNSS}\textup, is quantum for $X$.
\end{mytheo}

Diffeologists can regard Definition \eqref{QR1} as a special case of \eqref{QR2}, for they know that the base of a Kostant-Souriau line bundle $L\to X$ is always a coadjoint orbit of $\Aut(L)$ in the diffeological sense \cite[{}4.3b]{Souriau:1988}. Repeating the proof of \eqref{geomcrit1} we can again recast the definition in more geometrical fashion, as follows.

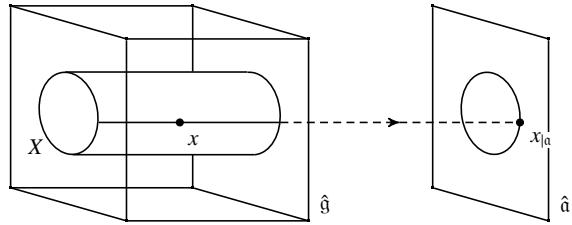
\begin{figure}[b!]
  \centering
  \vspace{1ex}
  \begin{tikzpicture}[scale=1.1,inner sep=0pt, outer sep=0pt]
     \node [fill=black] (A) at (0,0) {};
     \node [fill=black] (B) at (0,2.2) {};
     \node [fill=black] (C) at (1.4,1.8) {};
     \node [fill=black] (D) at (1.4,-.4) {};
     \node [fill=black] (E) at (2.2,0) {};
     \node [fill=black] (F) at (2.2,2.2) {};
     \node [fill=black] (G) at (3.6,1.8) {};
     \node [fill=black] (H) at (3.6,-.4) {};
     \node at (3.8,-.2) {$\hat\LG$};
     \node [fill=black] (I) at (5.1,0) {};
     \node [fill=black] (J) at (5.1,2.2) {};
     \node [fill=black] (K) at (6.5,1.8) {};
     \node [fill=black] (L) at (6.5,-.4) {};
     \node at (6.7,-.2) {$\hat\LA$};
     \node at (6.8,-.2) {};
     \draw [semithick] (A) -- (B) -- (C) -- (D) -- (A);
     \draw [semithick] (F) -- (G) -- (H) -- (E);
     \draw [semithick] (I) -- (J) -- (K) -- (L) -- (I);
     \draw [semithick] (A) -- (E);
     \draw [semithick] (B) -- (F);
     \draw [semithick] (C) -- (G);
     \draw [semithick] (D) -- (H);
     \draw [semithick,rotate=100] ($(A)!.5!(C)$) ellipse [x radius=.5cm,y radius=.35cm];
     \draw [semithick,rotate=100] ($(I)!.5!(K)$) ellipse [x radius=.5cm,y radius=.35cm];
     \draw [semithick] (0.66,1.4) -- (2.86,1.4);
     \draw [semithick] (0.75,0.4) -- (2.95,0.4);
     \draw [semithick,rotate around={95:(2.95,0.4)}] (2.95,0.4) arc (180:360:.5cm and .35cm);
     \draw [semithick] (E) -- ($(E)!.18!(F)$);
     \draw [semithick] ($(E)!.64!(F)$) -- (F);
     \node (M) at (1.05,0.79) {};
     \node (N) at (3.26,0.79) {};
     \draw [semithick] (M) -- (N);
     \draw [semithick, densely dashed,->,>=stealth'] ($(M)!1!(N)$) -- ($(M)!1.65!(N)$);
     \draw [semithick, densely dashed] ($(M)!1.65!(N)$) -- ($(M)!2.33!(N)$);
     \filldraw ($(M)!2.31!(N)$) circle (1.2pt);
     \filldraw ($(M)!.45!(N)$) circle (1.2pt);
     \node at (.3,.5) {$X$};
     \node at (2.2,.58) {$x$};
     \node [fill=white] at (6.41,.58) {$x\suba$};
  \end{tikzpicture}
  \vspace{1ex}
  \caption{Projection of a coadjoint orbit $X$ of $G$ to the dual of an abelian subalgebra $\LA\subset\LG$.}
  \label{fig:projection}
\end{figure}

\begin{mytheo}[\cite{Ziegler:1996b}; Fig.~\ref{fig:projection}]
   \label{geomcrit2}
   A state $m$ of $G$ is quantum for $X$ if and only if for each abelian subalgebra $\LA$ of $\LG$\textup, the state $m\comp\exp\suba$ of $\LA$ has its spectral measure concentrated on the closure $\b X\suba$ of $X\suba$ in $\hat\LA$.
\end{mytheo}

Here $\suba$ means restriction to $\LA$, and as before $\hat\LA$ denotes the (compact) character group of the \emph{discrete} additive group $\LA$. This densely contains the group of all \emph{continuous} characters, which we may and will identify with $\LA^*$ by letting $u\in\LA^*$ stand for the character $\E^{\I\<u,\,\cdot\,\>}$ of $\LA$. Likewise we define $\hat\LG$ and regard $\LG^*$ as a dense subgroup; in doing so we must be careful to distinguish between usual closure in $\LG^*$ and closure in $\hat\LG$, which we denote by $X\to \b X$ for \emph{\textbf{Bohr closure}}. Finally we remark that the notation $\b X\suba$ is unambiguous, i.e.~we have $(\b X)\suba=\b(X\suba)$: the projection of $\b X$ lies in the closure of $X\suba$ by continuity; moreover it is compact and so contains this~closure.

Now the point of \eqref{geomcrit2} is that the effect of Bohr closure is quite drastic:

\begin{mytheo}[\cite{Howe:2014}] \  
   \label{Howe-Ziegler}
   \begin{itemize}
      \vspace{-1ex}
   	\item[\textup{(a)}] If $G$ is noncompact simple\textup, any nonzero coadjoint orbit is Bohr dense in $\hat\LG$.
   	\item[\textup{(b)}] If $G$ is connected nilpotent\textup, any coadjoint orbit has the same Bohr closure as its affine hull.
   \end{itemize}
\end{mytheo}

\begin{mycoro} \ 
   \label{all_are}
   \begin{itemize}
      \vspace{-1ex}
   	\item[\textup{(a)}] If $G$ is noncompact simple\textup, every unitary representation is quantum for every nonzero coadjoint orbit.
    	\item[\textup{(b)}] If $G$ is simply connected nilpotent\textup, a unitary representation is quantum for an orbit $X$ if and only if its restriction to $C_X = \exp\left(\bigl\{Z\in\LG : \<\cdot,Z\> \textrm{ is constant on } X\bigr\}\right)$ is the character  $\exp(Z)\mapsto\E^{\I\<X,Z\>}$ times the identity.
   \end{itemize}
\end{mycoro}

(Here of course $\<X,Z\>$ denotes the common value of $\<x,Z\>$ for all $x\in X$.)

\begin{proof}
	(a) is immediate from Theorems (\ref{geomcrit2}) and (\ref{Howe-Ziegler}a). To prove (b), let $\mathcal H$ be a unitary $G$-module, pick a unit vector  $\varphi\in\mathcal H$ and write $m(g) = (\varphi,g\varphi)$.
   
	Suppose $\mathcal H$ is quantum for $X$. If $\LA$ is any 1-dimensional subalgebra of $\LC_X$, then $X\suba$ consists of the single point $Z\mapsto\<X,Z\>$. So \eqref{geomcrit2} says that $m(\exp(Z)) = \E^{\I\<X,Z\>}$ for all $Z\in\LA$ and hence for all $Z\in\LC_X$. Since $\|g\varphi-m(g)\varphi\|^2 = 1 - |m(g)|^2$ this implies that $C_X$ acts by $\exp(Z)\varphi= \E^{\I\<X,Z\>}\varphi$, as claimed.
 
	Conversely, suppose that $C_X$ acts by this character. Let $\LA$ be any abelian subalgebra of $\LG$, and write $\iota:\LA\cap\LC_X\to\LA$ for the natural injection and $\iota^*:\LA^*\to(\LA\cap\LC_X)^*$ and $\hat\iota:\hat\LA\to(\LA\cap\LC_X)\hat{\phantom{\LA}}$ for the dual projections. The relation $m\comp\exp_{|\LA\cap\LC_X} = m\comp\exp\suba\comp\,\iota$ shows that the spectral measure of $m\comp\exp_{|\LA\cap\LC_X}$ is the image by $\hat\iota$ of that of $m\comp\exp\suba$. As the former is concentrated on the point $X_{|\LA\cap\LC_X}$ by hypothesis, it follows that the latter is concentrated on the preimage $\hat\iota\inv(X_{|\LA\cap\LC_X})$ of this point \cite[n$^{\textrm o}$ V.6.2, Cor.~2]{Bourbaki:1967b}. There remains to see that this preimage is precisely $\b X\suba$. This follows from the calculation
	\begin{equation}
	   \hat\iota\inv(X_{|\LA\cap\LC_X}) = 
	   \b\iota^{*-1}(X_{|\LA\cap\LC_X}) =
	   \b\Aff(X\suba) = 
	   \b\Aff(X)\suba =
	   \b X\suba
	\end{equation}
	where `Aff' stands for affine hull. Here the first equality is because both $\hat\iota\inv(X_{|\LA\cap\LC_X})$ and $\iota^{*-1}(X_{|\LA\cap\LC_X})$ are preimages of points, hence translates of closed subgroups. The second equality is because the affine hull of $X\suba$ is the intersection of all hyperplanes containing it. The third is because the affine hull of a projected set is the projection of its affine hull. And finally the fourth equality is Theorem (\ref{Howe-Ziegler}b).
\end{proof}
		
\begin{myremas}
   The results \eqref{all_are} were certainly unexpected by the author of Definition \eqref{QR2}. They are in sharp contrast with our findings in §2: while it was easy to find non-quantum representations of $\Aut(L)$, but unknown if a quantum one even exists (a question whose difficulty is probably on par with that of making sense of the Feynman integral), scaling our ambitions back to finding representations of \emph{Lie subgroups} has now produced the opposite situation, where quantum representations are in such rich supply that it may even be impossible (\ref{all_are}a) to find a non-quantum one! 
   Clearly this indicates that---whatever may be the case of Definition \eqref{QR1}---Definition \eqref{QR2} still needs to be refined.
   
   One way to do so is to keep our hopes up high in \eqref{QR1} and bet that asking for states that \emph{extend to} $\Aut(L)$ will provide the much-needed selection. (Note that extending a state is a very different proposition from extending the resulting representation \emph{in the same space}, as Van Hove was trying to do. The GNS module \eqref{GNSS} of an extended state is usually much bigger than that of the state's restriction to a subgroup.)
   
   A second, more conservative way is to lay the blame for \eqref{all_are} on the Bohr closure in \eqref{geomcrit2} as the obvious culprit, and just suppress this closure. (Here we note that compactifying $X$ is really a change at the \emph{classical} level: our quantum states have probability measures on $\b X$ rather than $X$ as their classical analogues. In fact Souriau's papers \cites{Souriau:1988}{Souriau:1990a}{Souriau:1992} contain also a theory of ``statistical states'' which boil down to just that, probability measures on $\b X$.) This path was explored in \cite{Ziegler:1996b} with mixed results: one does recover the ``orbit methods'' of Borel-Weil and Kirillov-Bernat as special cases, but only after adding one or two hypotheses which may seem \emph{ad hoc}. 
\end{myremas}

\section{Localized states}

In this paper we want to explore a third way---one that doesn't suppress the compactification of $X$ implicit in Definition \eqref{QR2}, but instead takes it seriously. Our investigation is motivated by the discovery, among quantum states, of objects that solve in some cases (albeit in a rather unexpected way), what A.~Weinstein \cite{Weinstein:1982} has called the \emph{fundamental quantization problem}: to attach (possibly distributional) ``wave functions'' to lagrangian submanifolds of $X$. It will turn out that these states not only exist, but can be uniquely characterized quite simply:

\begin{mydefi}
   \label{localized}
   Let $X$ be a coadjoint orbit of the Lie group $G$, and $Y$ a coadjoint orbit of a closed subgroup $H\subset G$, contained in $X_{|\LH}$. We say that a quantum state $m$ for $X$ is \emph{\textbf{localized at}} $Y\subset\LH^*$ if the restriction $m_{|H}$ is a quantum state for $Y$.
\end{mydefi}

We also think of this as meaning that the state is \emph{\textbf{localized on}} $\pi\inv(Y)$, where $\pi$ is the projection $X\to\LH^*$. We recall from \cite[Prop.~1.1]{Kazhdan:1978} that this set is generically a \emph{coisotropic} submanifold of $X$---hence at least half-dimensional, and suitable for constraining a system to.

We shall almost exclusively apply Definition \eqref{localized} to cases where $H$ is connected and $Y=\{y\}$ is a point-orbit. To be a quantum state for $\{y\}$ then means the following. 

\begin{mypropdefi}[Integral point-orbits]
   \label{point_orbits}
   Let $H$ be a connected Lie group and $\{y\}$ a point-orbit of $H$ in $\LH^*$. A quantum state $n$ of $H$ for $\{y\}$ exists if and only if $y$ is \textbf{integral} in the sense that $H$ admits a character $\chi$ with differential $\I y$. It is then unique and given by that character\textup, i.e.~$n(\exp(Z))$ equals
   \begin{equation}
      \label{character}
      \chi(\exp(Z))=\E^{\I\<y,Z\>}.
   \end{equation}
\end{mypropdefi}

\begin{proof}
   Since $y$ is an $H$-invariant point in $\LH^*$, we have $\<y,[Z,Z']\>=0$ for all $Z, Z'\in\LH$. Thus $\I y$ defines a Lie algebra homomorphism from $\LH$ to the abelian Lie algebra $\Lie u(1)=\I\RR$. This integrates into a character $\tilde\chi:\tilde H\to\mathrm U(1)$ of the simply connected covering group $\tilde H$ of $H$, which descends to $H$ if and only if $y$ is integral.
   
   Suppose that $n$ is a quantum state for $\{y\}$. For each line $\LA=\RR Z$ in $\LH$, Theorem \eqref{geomcrit2} says that $n\comp\exp\suba$ has its spectral measure concentrated on the point $\{y\suba\}$, hence must be given by $(n\comp\exp\suba)(Z)=\smash{\E^{\I\<y,Z\>}}$. Therefore $n$ must coincide with $\chi$.
\end{proof}

\begin{mycoro}
	\label{eigenvector}
   Suppose $H$ is a closed connected subgroup of the Lie group $G$\textup, and $\{y\}$ an integral point-orbit of $H$ in $\LH^*$ with resulting character $\chi$ \eqref{character}. Then a quantum state $m$ of $G$ is localized at $\{y\}\subset\LH^*$ if and only if the cyclic vector \textup{(\hyperlink{cyclic_vector}{A.7})} of the resulting \textup{GNS} module is an eigenvector of type $\chi$ under $H$.
\end{mycoro}

\begin{proof}
   Suppose $m$ is localized at $\{y\}$, i.e.~$m_{|H}$ is a quantum state for $\{y\}$. Then we have $m_{|H}=\chi$ by the previous Proposition, and \eqref{Weil} implies $m(hg)=\chi(h)m(g)$ for all $(g,h)\in G\times H$. Therefore the cyclic vector $\varphi=\overline m$ satisfies
   \begin{equation}
      (h\overline m)(g) = 
      \overline m(h\inv g) =
      \overline m(h\inv)\overline m(g) =
      \chi(h)\overline m(g),
   \end{equation}
   i.e. $h\varphi = \chi(h)\varphi$, as claimed. Conversely, suppose that this last relation holds. Then we have $m(h) = (\varphi, h\varphi) = (\varphi, \chi(h)\varphi) = \chi(h)$. So $m_{|H}=\chi$, which is to say that $m$ is localized at $\{y\}$.
\end{proof}

Definition \eqref{localized} will allow us to extract interesting objects from the generally unclassifiable maze \eqref{all_are} of all quantum representations. This is somewhat reminiscent of the representation theory of Lie algebras, where one can't in general describe the class of simple modules \cite{Benoist:1990a}, but where imposing the presence of eigenvectors produces a manageable classification problem \cite{Benoist:1990}.

\section{Nilpotent groups}

In this section we assume that $G$ is a connected, simply connected nilpotent Lie group. Then $\exp:\LG\to G$ is a diffeomorphism whose inverse we denote $\log:G\to\LG$. Moreover we fix a coadjoint orbit $X\subset\LG^*$ and a point $x\in X$, and we recall that a connected subgroup $H=\exp(\LH)$ of $G$ is called \emph{\textbf{subordinate to}} $x$ if, equivalently,
\begin{itemize}
   \item[(a)] $\E^{\I x\,\comp\,\log}{}_{|H}$ is character of $H$;
   \item[(b)] $x_{|\LH}$ is a point-orbit of $H$ in $\LH^*$;
   \item[(c)] $\<x,[\LH,\LH]\> = 0$.
\end{itemize}
Any subordinate subgroup has $\dim(G/H)\geqslant\frac12\dim(X)$; if this bound is attained then one calls $H$ a \emph{\textbf{polarization at}} $x$. Polarizations are maximal subordinate subgroups, but some maximal subordinate subgroups are not polarizations.

\begin{mytheo}
   \label{nilpotentthm}
	Let $H$ be maximal subordinate to $x\in X$. Then there is a unique quantum state for $X$ localized at $\{x_{|\LH}\}\subset\LH^*$\textup, namely
	\begin{equation}
	   \label{localized_nilpotent}
      m(g)= 
	   \begin{cases}
	      \ \ \E^{\I x\,\comp\,\log}(g) & \text{if $g\in H$\textup,}\\
	      \ \ 0 & \text{otherwise.}
	   \end{cases}
   \end{equation}
	The associated $\GNS{}$ representation \eqref{GNSS} is $\ind_H^G \E^{\I x\,\comp\,\log}{}_{|H}$\textup, where induction is in the sense of discrete groups.\footnote{Here and elsewhere we reserve the lower case `$\ind$' for discrete induction, as opposed to the usual `Ind' when $G$ already has another locally compact topology.}
\end{mytheo}

\begin{proof}
   The fact that $m$ must coincide with $\E^{\I x\,\comp\,\log}$ in $H$ is just \eqref{point_orbits}. To see that it must vanish outside $H$, we consider the sequence $H=G_0\subset G_1\subset G_2\dots$ where $G_{i+1}$ is the normalizer of $G_i$ in $G$. Since $G$ is nilpotent, the $G_i$ are connected and all equal to $G$ after finitely many steps \cite[Prop.~III.9.16]{Bourbaki:1972}; so it is enough to show inductively that $m$ vanishes in $G_{i+1}\smallsetminus G_i$ for all $i$.

   \emph{Case $i=0$.} Let $g\in G_1\smallsetminus H$. Applying Weil's inequality \eqref{Weil} twice, we get
   \begin{equation}
      \label{Weil_twice}
	   \E^{\I x\,\comp\,\log}(h)m(g)
	   =m(hg)
	   =m(gg\inv hg)
	   =m(g)\E^{\I x\,\comp\,\log}(g\inv hg)
   \end{equation}
   for all $h\in H$. Thus, if $m(g)$ was nonzero, $g$ would both normalize $H$ and stabilize its character $\E^{\I x\,\comp\,\log}{}_{|H}$. Since the normalizer and stabilizer in question are connected \cites{Bernat:1972}{Bourbaki:1972} it would follow that $Z=\log(g)$ normalizes $\LH$ and stabilizes $x_{|\LH}$. Putting $\LK=\LH\oplus\RR Z$, we would conclude that $\<x,[\LK,\LK]\>$ is zero. But then $K=\exp(\LK)$ would be subordinate to $x$, and so $H$ would not be maximal subordinate to $x$. This contradiction shows that $m(g)=0$.

   \emph{Case $i>0$.} Let $g\in G_{i+1}\smallsetminus G_i$. Then $g$ normalizes $G_i$ but not $H$, so we can fix an $h\in H$ such that $g\inv hg\in G_i\smallsetminus H$. Putting $g_n=h^ng$ it follows that $\smash{g_p\inv g_q^{\phantom1}}\in G_i\smallsetminus H$ whenever $p\ne q$. The induction hypothesis then shows that $m(\smash{g_p\inv g_q^{\phantom1}})=0$, which is to say that the $\d^{g_n}$ ($=1$ at $g_n$ and $0$ elsewhere) make an orthonormal set relative to the sesquilinear form \eqref{sesquilinear}. Therefore Bessel's inequality gives
   \begin{equation}
	   \sum_n|m(g_n)|^2=\sum_n|(\d^e,\d^{g_n})_m|^2\leqslant(\d^e,\d^e)_m=1.
   \end{equation}
   Now this forces $m(g)=0$, because we have $|m(g_n)|=|\E^{\I x\,\comp\,\log}(h^n)m(g)|=|m(g)|$ for all $n$. Finally the last assertion of the Theorem is a special case of \eqref{Blattner}, and the fact that the state \eqref{localized_nilpotent} is indeed quantum for $X$ will result from (\ref{polarization_dependance}c) below, because maximal subordinate subgroups always contain $C_X$ \textup{(\ref{all_are}b)}.
\end{proof}

The representations
\begin{equation}
	\I(x,H)=\ind_H^G \E^{\I x\,\comp\,\log}{}_{|H}
\end{equation}
 found in \eqref{nilpotentthm} make sense whenever $H$ is subordinate to $x$, and are closely \mbox{analogous} to the representations $\mathrm I(x,H)=\Ind_H^G \E^{\I x\,\comp\,\log}{}_{|H}$ fundamental in Kirillov's theory \cite{Kirillov:1962}. These enjoy, we recall, the following key properties:
\begin{enumerate}
   \item[(a)] $\mathrm I(x,H)$ is irreducible if and only if $H$ is a \emph{polarization} at $x$.
   \item[(b)] $\mathrm I(x,H)$ and $\mathrm I(x,K)$ are \emph{equivalent} if $H$ and $K$ are any two polarizations at $x$.
\end{enumerate}
In sharp contrast to this, we shall prove:

\begin{mytheo} \ 
   \label{polarization_dependance}
   \vspace{-1ex}
	\begin{enumerate}
	   \item[\textup{(a)}] $\I(x,H)$ is irreducible if and only if $H$ is maximal subordinate to $x$.
	   \item[\textup{(b)}] $\I(x,H)$ and $\I(x,K)$ are inequivalent whenever $H$ and $K$ are any two different polarizations at $x$.
	   \item[\textup{(c)}] $\I(x,H)$ is quantum for $X$ if and only if $H$ contains $C_X$ \textup{(\ref{all_are}b)}.
	\end{enumerate}
\end{mytheo}

\begin{proof}
   (a): Suppose that $H$ is subordinate to $x$ but not maximally so, i.e., $H$ is strictly contained in another subordinate subgroup $K$. Since $K$ is nilpotent, the normalizer $N$ of $H$ in $K$ contains $H$ strictly \cite[Prop.~III.9.16]{Bourbaki:1972}. Now, given $s\in N\smallsetminus H$, one verifies readily that $(Jf)(g)=f(gs)$ defines a unitary intertwining operator $J:\I(x,H)\to\I(x,H)$ which is not scalar since $(m_e,Jm_e)=0$ (\ref{repro_kernel}, \ref{chi^bullet}). So $\I(x,H)$ is reducible.
   
   Conversely, suppose $\I(x,H)$ is reducible. Then some double coset $D=HgH$, other than $H$, must satisfy the Mackey-Shoda conditions  \eqref{Mackey-Shoda} with $\chi=\eta=\E^{\I x\,\comp\,\log}{}_{|H}$. But then $g$ must normalize $H$: indeed, if some $h\in H$ were outside $gHg\inv$, so would be $h^n$ for all $n\ne0$; so we would have $h^pgH\ne h^qgH$ whenever $p\ne q$, and so $D/H$ would be infinite, contradicting (\ref{Mackey-Shoda}b). Thus $g$ normalizes $H$ and stabilizes $\E^{\I x\,\comp\,\log}{}_{|H}$ (\ref{Mackey-Shoda}a), and we conclude just as in the proof of \eqref{nilpotentthm}(case $i=0$) that $H$ is not maximal subordinate to $x$.
      
   (b): Let $H$ and $K$ be polarizations at $x$, and suppose there is a double coset $D=HgK$ satisfying the conditions of \eqref{Mackey-Shoda} with $\chi=\E^{\I x\,\comp\,\log}{}_{|H}$, $\eta=\E^{\I x\,\comp\,\log}{}_{|K}$. As above, it follows that $H=gKg\inv$ and $\chi(h)=\eta(g\inv hg)$ for all $h\in H$. Thus we have $\E^{\I\<x-g(x),\LH\>}= 1$, or in other words, $g(x)\in x+\LH^\perp= H(x)$ \cite[pp.\,69--70]{Bernat:1972}. Since $H$ contains the stabilizer $G_x$, this forces $g\in H$ and hence $K=H=D$. Thus, \eqref{Mackey-Shoda} says that $\Hom_G(\I(x,H),\I(x,K))$ has dimension 1 if $H=K$, and 0 otherwise.
   
   (c): We know from \eqref{Blattner} that $\I(x,H)=\GNS{m}$ where $m$ is the state \eqref{localized_nilpotent}. By \eqref{GNS_quantum}, this module is quantum for $X$ if and only if $m$ is. By (\ref{all_are}b), that is true if and only if \eqref{localized_nilpotent} coincides with $\E^{\I x\,\comp\,\log}$ on $C_X$, which is to say that $C_X$ lies in $H$.
\end{proof}

\begin{myexam}[Heisenberg's orbit]
   The results (\ref{nilpotentthm}, \ref{polarization_dependance}) are already instructive in the simplest case of the group \eqref{Heisenberg} with Lie algebra
   \begin{equation}
      \label{Lie(Heisenberg)}
      \LG=\left\{Z=
      \begin{pmatrix}
         \,0 & \,\beta\, & \alpha\,\\
           & 0     & \gamma\,\\
           &       & 0\,
      \end{pmatrix}
      : \alpha, \beta, \gamma\in\RR
      \right\}.
   \end{equation}
   We consider the coadjoint orbit $X$ of the linear form $Z\mapsto -\alpha $. It is isomorphic~to $(\RR^2,dp\wedge dq)$ under the map $\Phi$ given by
   $
   \<\Phi(p,q),Z\>=\,
   \smash{\left|\begin{smallmatrix} p&\,q\\\beta&\gamma\end{smallmatrix}\right|}
   -\alpha
   $. By (\ref{all_are}b), a state $m$ of $G$ is quantum for $X$ if and only if it restricts to the character $\E^{-\I a}$ of the center (\ref{Heisenberg_modules}d). Its statistical interpretation then gives (among others) the variables $p$ and $q$ probability distributions $\mu$ and $\nu$ defined by
   \begin{flalign}
      \label{munu}
      &
      \textstyle
      \int_{\b\RR}
      \E^{\I(p,\gamma)}
      \,d\mu(p)
      =
      m\comp\exp_{|\LC}
      {\small
      \left(
      \begin{smallmatrix}
         0 & 0 & 0\\
           & 0 & \gamma\\
           &   & 0
      \end{smallmatrix}
      \right),
      }
      &
      \textstyle
      \int_{\b\RR}
      \E^{-\I(q,\hspace{1pt}\beta)}
      \,d\nu(q)
      =
      m\comp\exp_{|\LB}
      {\small
      \left(
      \begin{smallmatrix}
         0 & \beta & 0\\
           & 0 & 0\\
           &   & 0
      \end{smallmatrix}
      \right).
      }
      \hspace{2ex}
   \end{flalign}
   Here we write elements of the Bohr compactification $\b\RR$ as (possibly discontinuous) homomorphisms $(p,\cdot)$ and $(q,\cdot):\RR\to\RR/2\pi\ZZ$, and $\LB$, $\LC$ are the one-dimensional subalgebras of matrices of the indicated form. Choosing $x=\Phi(k,\ell)$ say, we have 
   \begin{equation}
      \E^{\I x\,\comp\,\log}(g)=\E^{-\I a}\E^{\I bc/2}\E^{\I(kc - \ell b)}
   \end{equation}
   and the maximal subordinate subgroups to $x$ are the polarizations $H_t$ ($t\in\RR\cup\infty$) listed in Table \ref{table}.

   \renewcommand\arraystretch{2.0}
   \setlength\arraycolsep{1.5pt}
   \begin{table}[h]
      \centering
      \begin{equation}
         \label{Heisenberg_modules}
         {\small
         \begin{array}{|c|c|c|c|}
            \hline
            &
            \textrm{Representation} & 
            \textrm{acts on $\ell^2$ functions} & 
            \textrm{by}\\\hline\hline
            \textrm{(a)} &
            \I\left(x,H_\infty=\left\{\left(
            \begin{smallmatrix}
               1 & 0 & a\\
                 & 1 & c\\
                 &   & 1
            \end{smallmatrix}
            \right)\right\}\right) &
            \phi(p)= f\left(
            \begin{smallmatrix}
               1 & p-k & 0\\
                 & 1 & 0\\
                 &   & 1
            \end{smallmatrix}
            \right) &
            (g\phi)(p) = \E^{-\I a}\E^{\I pc}\phi(p-b)\\[1ex]\hline
            \textrm{(b)} &
            \I\left(x,H_0=\left\{\left(
            \begin{smallmatrix}
               1 & b & a\\
                 & 1 & 0\\
                 &   & 1
            \end{smallmatrix}
            \right)\right\}\right) &
            \psi(q)= f\left(
            \begin{smallmatrix}
               1 & 0 & 0\\
                 & 1 & q-\ell\\
                 &   & 1
            \end{smallmatrix}
            \right) &
            (g\psi)(q) = \E^{-\I a}\E^{-\I b(q-c)}\psi(q-c)\\[1ex]\hline
            \textrm{(c)} &
            \I\left(x,H_t=\left\{\left(
            \begin{smallmatrix}
               1 & b & a\\
                 & 1 & -bt\\
                 &   & 1
            \end{smallmatrix}
            \right)\right\}\right) &
            \psi(r)= f\left(
            \begin{smallmatrix}
               1 & 0 & 0\\
                 & 1 & r-\ell-kt\\
                 &   & 1
            \end{smallmatrix}
            \right) &
            (g\psi)(r) = \E^{-\I a}\E^{-\I b(r-c)}\E^{\I b^2t/2}\psi(r-c-bt)\\[1ex]\hline
            \textrm{(d)} &
            \I\left(x,C_X=\left\{\left(
            \begin{smallmatrix}
               1 & 0 & a\\
                 & 1 & 0\\
                 &   & 1
            \end{smallmatrix}
            \right)\right\}\right) &
            \varphi(p,q)= f\left(
               \begin{smallmatrix}
                  1 & p & 0\\
                    & 1 & q\\
                    &   & 1
               \end{smallmatrix}
               \right) & 
               (g\varphi)(p,q)=\E^{-\I a}\E^{-\I b(q-c)}\varphi(p-b,q-c)\\[1ex]\hline
         \end{array}
         }
      \end{equation}
      \caption{The representations $\I(x,H)$ attached to the subordinate subgroups $H_t$ $(t\in\RR\cup\infty)$ and~$C_X$. While each acts nominally in sections of $G\times_H\CC\to G/H$, i.e.~on equivariant functions $f:G\to\CC$ (\protect\hyperlink{G-action}{A.6}, \ref{Blattner}a), the middle column trivializes this bundle to realize the representation in $\ell^2(G/H)$.}
      \label{table}
   \end{table}

   (a): A state localized at $\{x_{|\LH_\infty}\}\subset\LH_\infty^*$ is one in which $p$ is certainly $k$. Theorem \eqref{nilpotentthm} asserts that there is a unique such state, which is discontinuous: $m(g)=\E^{-\I a}\d^b\sub0\E^{\I kc}$ (Kronecker's delta). Its statistical interpretation \eqref{munu} reads
   \begin{flalign}
      &
      \hspace{0.5ex}
      \textstyle
      \int_{\b\RR}
      \E^{\I(p,\gamma)}
      \,d\mu(p)
      = \E^{\I k\gamma},
      &
      \textstyle
      \int_{\b\RR}
      \E^{-\I(q,\hspace{1pt}\beta)}
      \,d\nu(q)
      =
      \left\{
      \begin{smallmatrix}
         \rlap{\small1\quad if $\beta=0$}\\
         \rlap{\small0\quad otherwise,}
      \end{smallmatrix}
      \right.
      \hspace{15ex}
   \end{flalign}
   i.e. while $\mu$ is Dirac measure at $k$ (as desired), $\nu$ is Haar measure on $\b\RR$ \eqref{1_H}. So Theorem \eqref{nilpotentthm} entails a version of Heisenberg's principle: \emph{$p$ may be certain\textup, but then $q$ is necessarily equidistributed on the whole line}. 
   
   The GNS representation $\I(x,H_\infty)$ obtained from $m$  \eqref{nilpotentthm} was apparently first considered (as representation of a certain C${}^*$-algebra) in the papers \cites{Beaume:1974}{Emch:1981}. It acts in $\ell^2(\RR)$ by the very same action (\ref{Heisenberg_modules}a) by which the Schrödinger representation $\mathrm I(x,H_\infty)$ acts in $\mathrm L^2(\RR)$. We know from (\ref{polarization_dependance}a) that it is irreducible, and from \eqref{nilpotentthm} that its cyclic vector $\phi(p)=\d^p_k$ (obtained by taking $f=\overline m$ in (\ref{Heisenberg_modules}a); cf.~(\hyperlink{cyclic_vector}{A.7})) is an eigenvector of the ``translation'' subgroup $\exp(\LC)$---befitting the fact that the resulting measure $\nu$ is translation-invariant.
   
   (b): A state localized at $\{x_{|\LH_0}\}\subset\LH_0^*$ is one in which $q$ is certainly $\ell$. Again \eqref{nilpotentthm} provides the unique such state: $m(g) = \E^{-\I a}\E^{-\I \ell b}\d^c\sub0$, with statistical interpretation
   \begin{flalign}
      &
      \hspace{0.5ex}
      \textstyle
      \int_{\b\RR}
      \E^{\I(p,\gamma)}
      \,d\mu(p)
      =
      \left\{
      \begin{smallmatrix}
         \rlap{\small1\quad if $\gamma=0$}\\
         \rlap{\small0\quad otherwise,}
      \end{smallmatrix}
      \right.
      &
      \textstyle
      \int_{\b\RR}
      \E^{-\I(q,\hspace{1pt}\beta)}
      \,d\nu(q)
      = \E^{-\I \ell\beta},
      \hspace{12.2ex}
   \end{flalign}
   i.e.~$\mu$ is Haar measure on $\b\RR$ while $\nu$ is Dirac measure at $\ell$. The resulting representation (\ref{Heisenberg_modules}b) is sometimes called the \emph{polymer representation}, after \cite[§III.B]{Ashtekar:2003}. Although related to (\ref{Heisenberg_modules}a) by an automorphism of $G$, it is inequivalent to it (\ref{polarization_dependance}b). Its cyclic vector, $\psi(q) = \d^q_\ell$, is now an eigenvector of the ``boost'' subgroup $\exp(\LB)$.
   
   (c): More generally, a state localized at $\{x_{|\LH_t}\}\subset\LH_t^*$ is one in which $q+pt$ is certainly $\ell+kt$. To further illustrate why the resulting modules (\ref{Heisenberg_modules}c) are inequivalent for different values of $t$ (\ref{polarization_dependance}b), we map their space $\ell^2(\RR)$ to $\mathrm L^2(\b\RR)$ by the Fourier transform $\hat\psi(p)=\sum\E^{\I(p,r)}\psi(r)$ and compute the transported actions, obtaining
   \begin{equation}
	   \label{cocycle}
	   (g\hat\psi)(p)= \E^{-\I a}\E^{\I(p,c)}\E^{\I((p,bt) - \frac12b^2t)}\hat\psi(p-b).
	\end{equation}
	In $\mathrm L^2(\RR)$, these actions are all unitarily equivalent to each other (and to (\ref{Heisenberg_modules}a)), because the factor $\E^{\I(pbt - \frac12b^2t)}$ is the coboundary, $u(p-b)/u(p)$, of $u(p)=\E^{-\I p^2t/2}$. But in $\mathrm L^2(\b\RR)$ that is no longer the case, because $u$ is not almost periodic.
	
	(d): Finally (and unrelated to localization), (\ref{polarization_dependance}c) lets us induce from the center $C_X$ itself, using $m(g)=\E^{-\I a}\d^b\sub0\d^c\sub0$. The resulting module (\ref{Heisenberg_modules}d) is simply an $\ell^2$ version of the prequantization representation \eqref{reducible}. Like the latter, it is reducible (\ref{polarization_dependance}a) (and in fact finite type II \cite[{}Thm~11]{Kleppner:1962}); as such it would have been rejected by Van Hove, but Definition \eqref{QR2} welcomes it.
\end{myexam}

\begin{myrema}
	Another extant argument to discard (\ref{Heisenberg_modules}d) (or \eqref{reducible}) is that it \emph{``would violate the uncertainty principle since square integrable sections of $L$ can have arbitrarily small support''} \cite[p.\,7]{Sniatycki:1980}. This however is based on a misinterpretation of $\varphi(p,q)$, whose square modulus should not be regarded as a probability density in phase space. For example if $\varphi$ is the characteristic function of the origin, then the state $(\varphi,g\varphi)$ is our $m(g)=\E^{-\I a}\d^b\sub0\d^c\sub0$, whose statistical interpretation \eqref{munu} reads
	   \begin{flalign}
	      &
	      \hspace{0.5ex}
	      \textstyle
	      \int_{\b\RR}
	      \E^{\I(p,\gamma)}
	      \,d\mu(p)
	      =
	      \left\{
	      \begin{smallmatrix}
	         \rlap{\small1\quad if $\gamma=0$}\\
	         \rlap{\small0\quad otherwise,}
	      \end{smallmatrix}
	      \right.
	      &
	      \textstyle
	      \int_{\b\RR}
	      \E^{-\I(q,\hspace{1pt}\beta)}
	      \,d\nu(q)
	      =
	      \left\{
	      \begin{smallmatrix}
	         \rlap{\small1\quad if $\beta=0$}\\
	         \rlap{\small0\quad otherwise.}
	      \end{smallmatrix}
	      \right.
	      \hspace{15ex}
	   \end{flalign}
		So \emph{both $\mu$ and $\nu$} are Haar measure on $\b\RR$, and far from being concentrated at $0$, $p$ and $q$ are both equidistributed on the whole line.\footnote{One can also reason purely in the $\mathrm L^2$ version: although the function $\varphi_\epsilon(p,q) = {\scriptstyle\surd(2\pi\epsilon)\inv}\E^{-(p^2+q^2)/4\epsilon}$ ``shrinks to the origin'' as $\epsilon\to0$, one computes without trouble that the resulting state $(\varphi_\epsilon,g\varphi_\epsilon)$ (which incidentally, tends pointwise to $m$) assigns to $p$ and $q$ probability distributions whose product of variances, $\varDelta p\varDelta q= \frac12\surd\bigl(1+\frac1{4\epsilon^2}\bigr)$, tends not to zero but to infinity.}
\end{myrema}

\begin{myexam}[Bargmann's orbit]
   The effects of Bohr closure in the previous example were still rather mild, insofar as $X$ was equal to its affine hull (cf.~(\ref{Howe-Ziegler}b)). So we move on to the next simplest example, where $G$ (resp.~$\LG$) consists of all real matrices of the form
   \begin{equation}
      g=
      \begin{pmatrix}
         \,1\, & b & \frac12b^2 & a\,\\
           & 1 & b          & c\,\\
           &   & 1          & e\,\\
           &   &            & 1\,
      \end{pmatrix},
      \qquad\text{resp.}\qquad
      Z=
      \begin{pmatrix}
         \,0\, & \,\beta\,     & 0 & \,\alpha\,\\
           & 0 & \,\beta\, & \,\gamma\,\\
           &   & 0     & \,\varepsilon\,\\
           &   &       & \,0\,
      \end{pmatrix}.
   \end{equation}
   Forgetting the first row and column yields the Galilei group of space-time transformations 
   \begin{equation}
      g
      \begin{pmatrix}\,r\,\\t\\1\end{pmatrix}
      =
      \begin{pmatrix}\,1 & \,b\, & c\,\\& 1 & e\,\\&&1\,\end{pmatrix}
      \begin{pmatrix}\,r\,\\t\\1\end{pmatrix}
      =
      \begin{pmatrix}\,r+bt+c\,\\t+e\\1\end{pmatrix}
   \end{equation}
   of which $G$ is a central extension. We denote elements of $\LG^*$ as 4-tuples $(M,p,q,E)$, paired to $\LG$ by
   $
   \<x,Z\>=\,
   \smash{\left|\begin{smallmatrix} p&q\\\beta&\gamma\end{smallmatrix}\right|}
   -E\varepsilon - M\alpha
   $,
   and we consider the orbit of $(1,0,0,0)$. \linebreak It is isomorphic to  $(\RR^2,dp\wedge dq)$ under the map $\Phi$:
   \begin{equation}
      \label{paraboloid}
      \Phi(p,q) = \bigl(1,p,q,\tfrac12p^2\bigr).
   \end{equation}
   Theorem (\ref{all_are}b) says that a state $m$ of $G$ is quantum for $X$ if and only if it restricts to the character $\E^{-\I a}$ of the center $C_X=\{g: b=c=e=0\}$. Its statistical interpretation then assigns to the variables $(p,E)$ and $r:=q+pt$ ($t\in\RR$ a fixed parameter) proba\-bil\-ity distributions $\mu$ and $\nu_t$ defined by
   \begin{gather}
      \label{mu}
      \int_{\b\RR^2}\E^{\I[(p,\gamma)-(E,\varepsilon)]}d\mu(p,E)
      =
      m\comp\exp_{|\LC_{\phantom{t}}}
      \left(
      \begin{smallmatrix}
         0 & \,0\, & \,0\, & \,0\\
           & \,0\, & \,0\, & \,\gamma\\
           &       & \,0\, & \,\varepsilon\\
           &       &       & \,0
      \end{smallmatrix}
      \right),
      \\
      \label{nu_t}
      \int_{\b\RR}\E^{-\I(r,\,\beta)}d\nu_t(r)
      =
      m\comp\exp_{|\LB_t}
      \left(
      \begin{smallmatrix}
         0 & \,\beta & \,0     & 0\\
           & \,0     & \,\beta & -\beta t\\
           &         & \,0     & 0\\
           &         &         & 0
      \end{smallmatrix}
      \right),
   \end{gather}
   where $\LC$ and $\LB_t$ are the abelian subalgebras of matrices of the indicated form. Adding the center to $\LC$ and $\LB_t$ and exponentiating produces (abelian) subgroups $H_\infty$ and $H_t$ which turn out to be exactly all maximal subordinate subgroups to any $x=\Phi(k,\ell)$. Of these only $H_\infty$ is a polarization; the others are all conjugate under the stabilizer of $x$ in $G$, so it will suffice to specialize our results to $H_\infty$ and $H_0$ (Fig.~\ref{fig:Bargmann}).
\begin{figure}[t!]
  \centering
  \begin{tikzpicture}[scale=0.68,inner sep=0pt, outer sep=0pt]
     \def\zero{\left\{\left(\begin{smallmatrix}
        0 & \beta     & 0 & \alpha\\[-1pt]
          & 0 & \beta & 0\\[-1pt]
          &   & 0     & 0\\
          &   &       & 0
     \end{smallmatrix}\right)\right\}^*}
     \def\infinity{\left\{\left(\begin{smallmatrix}
        0 & 0 & 0 & \alpha\\
          & 0 & 0 & \gamma\\
          &   & 0 & \varepsilon\\
          &   &   & 0
     \end{smallmatrix}\right)\right\}^*}
     \node (S) at (2,.5) {};
     \node (T) at (5,0) {};
     \node [fill=black] (U) at (0,0) {};
     \node [fill=black] (V) at (3,-.5) {};
     \node [fill=black] (W) at (4,1) {};
     \node [fill=black] (X) at (7,.5) {};
     \node [fill=black] (Y) at (10,2.5) {};
     \node [fill=black] (Z) at (13,2) {};
     \draw [thick, name path=projected parabola] (Y) parabola[parabola height=-1cm] (Z);
     \draw [thick] (W) parabola[parabola height=-1cm] (X);
     \foreach \i in {0.52,0.54,...,0.88}
        \draw [color=white] ($(U)!\i!(W)$) parabola[parabola height=-1cm] ($(V)!\i!(X)$);
     \draw [thick, color=Cyan] (S) parabola[parabola height=-1cm] (T);
     \foreach \i in {0.02,0.04,...,0.48}
        \draw [color=white] ($(U)!\i!(W)$) parabola[parabola height=-1cm] ($(V)!\i!(X)$);
     \draw [thick] (U) parabola[parabola height=-1cm] (V);
     \node [fill=black] (A) at (0,-2.5) {};
     \node [fill=black] (B) at (0,1.5) {};
     \node [fill=black] (C) at (3,1) {};
     \node [fill=black] (D) at (3,-3) {};
     \draw [semithick] (A) -- (B) -- (C) -- (D) -- (A);
     \node [fill=black] (E) at (4,-1.5) {};
     \node [fill=black] (F) at (4,2.5) {};
     \node [fill=black] (G) at (7,2) {};
     \node [fill=black] (H) at (7,-2) {};
     \node (GG) at (7.3,-2) {$\ \LG^*$}; 
     \draw [opacity=0, name path=back corner] (E) -- (F);
     \draw [semithick] (F) -- (G) -- (H) -- (E);
     \node [fill=black] (I) at (10,0) {};
     \node [fill=black] (J) at (10,4) {};
     \node [fill=black] (K) at (13,3.5) {};
     \node [fill=black] (L) at (13,-.5) {};
     \node (FF) at (14.9,-.5) {$\LH_\infty^*=\infinity$}; 
     \draw [semithick] (I) -- (J) -- (K) -- (L) -- (I);
     \draw [semithick] (A) -- (E);
     \draw [semithick] (B) -- (F);
     \draw [semithick] (C) -- (G);
     \draw [semithick] (D) -- (H);
     \node (Q) at (1.75,-1.23) {};
     \node (R) at (5.75,-.23) {};
     \draw [thick,name path=paraboloid bottom] (Q) -- (R);
     \draw [thick] (U) -- (W);
     \draw [thick] (V) -- (X);
     \path [name intersections={of=back corner and paraboloid bottom}];
     \coordinate (M) at (intersection-1);
     \draw [semithick] (E) -- (M);
     \draw [semithick] (W) -- (F);
     \node [fill=black] (O) at (0.37,-0.55) {};
     \node [fill=black] (P) at (4.37,0.45) {};
     \draw [thick, color=Magenta] (O) -- (P);
     \draw [->,>=stealth',densely dashed,semithick] ($(O)!1!(P)$) -- ($(O)!1.75!(P)$);
     \draw [densely dashed,semithick] ($(O)!1.75!(P)$) -- ($(O)!2.5!(P)$);
     \filldraw [Magenta] ($(O)!2.5!(P)$) circle (2.5pt);
     \node [above right] (CC) at ($(O)!2.5!(P)$) {\ $(p,E)$};
     \node (AA) at (1.5,-4.75) {};
     \node (BB) at (5.5,-3.75) {};
     \node (EE) at (7.4,-3.75) {$\LH_0^*=\zero$};
     \draw [semithick] (AA) -- (BB);
     \draw [->,>=stealth',densely dashed,semithick] (3.5,-0.8) -- (3.5,-2.525);
     \draw [densely dashed,semithick] (3.5,-2.525) -- (3.5,-4.25);
     \filldraw [Cyan] (3.5,-4.25) circle (2.5pt);
     \node [below right] (DD) at (3.5,-4.25) {$\ \,q$};
     \node (GG) at (14,2.1) {$E=\frac12p^2$}; 
     \node [fill=black] (V) at (3,-.5) {};
  \end{tikzpicture}
  \caption{Projection of Bargmann's orbit \eqref{paraboloid} to the duals of  abelian subalgebras $\LH_0$ and $\LH_\infty$.}
  \label{fig:Bargmann}
\end{figure}
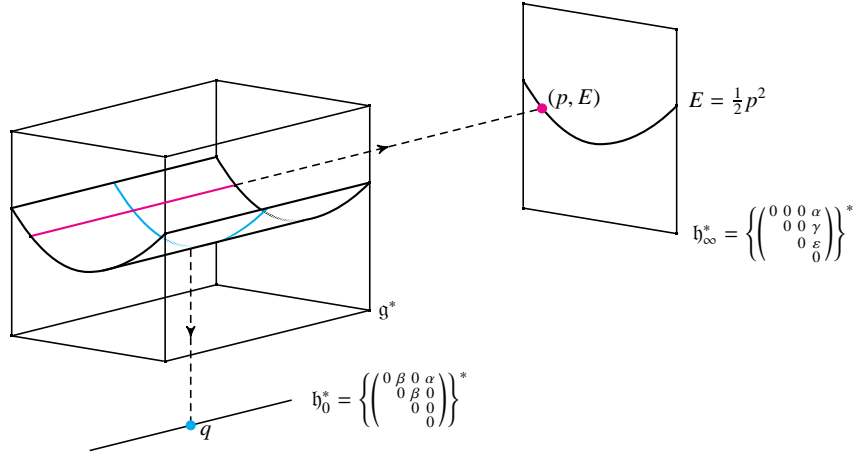

(a): A state localized at $\{x_{|\LH_\infty}\}\subset\LH_\infty^*$ is one in which $(p,E)$ is certainly $(k,\frac12k^2)$. Theorem \eqref{nilpotentthm} says that the unique such state is $m(g) = \E^{-\I a}\smash{\d^b\sub0\E^{\I(kc-\frac12k^2e)}}$. Computing as in (\ref{Heisenberg_modules}a), we find that the resulting representation $\I(x,H_\infty)$ acts in $\ell^2(\RR)$ by
\begin{equation}
   \label{Bargmann_representation}
   (g\phi)(p)=\E^{-\I a}\E^{\I(pc-\frac12p^2e)}\phi(p-b),
\end{equation}
with statistical interpretation as follows: in the state $(\phi,g\phi)$, the variable $p$ is distributed according to $|\phi(p)|^2$ times counting mesure on $\RR$, the pair $(p,E)$ according to the image of that measure under $p\mapsto(p,\frac12p^2)$, and the variable $r$ \eqref{nu_t} according to $\nu_t = |\psi(\begin{smallmatrix}r\\t\end{smallmatrix})|^2$ times Haar measure on $\b\RR$, where $\psi(\begin{smallmatrix}r\\t\end{smallmatrix}) = \sum_p\smash{\E^{-\I\{(r,p)-\frac12p^2t\}}}\phi(p)$. We note that the action of $G$ transported to these latter functions writes
\begin{equation}
   \label{on_solutions}
   (g\psi)(\begin{smallmatrix}r\\t\end{smallmatrix})
   =\E^{-\I a}\E^{-\I\{(r-c,\,b)-\frac12b^2(t-e)\}}
   \psi\bigl(\begin{smallmatrix}r-c-b(t-e)\\t-e\end{smallmatrix}\bigr),
\end{equation}
and that their restrictions to $r\in\RR\subset\b\RR$ constitute a \emph{non-standard Hilbert space of (almost-periodic) solutions} of  the Schrödinger equation $\I\partial_t\psi = \frac12\partial_r^2\psi$, with norm $\|\psi\|^2$ the Bohr mean of $|\psi(\begin{smallmatrix}\cdot\\t\end{smallmatrix})|^2$ (independent of $t$) and cyclic vector $\psi(\begin{smallmatrix}r\\t\end{smallmatrix})=\smash{\E^{-\I(kr - \frac12k^2t)}}$ (a ``plane wave'' \cite[§30]{Dirac:1930}). For comparison, the \emph{standard} solution space consists of transforms ${\scriptstyle\surd(2\pi)\inv}\!\!\int\smash{\E^{-\I(rp-\frac12p^2t)}}\phi(p)\,dp$ where $\phi\in\mathrm I(x,H_\infty)=\mathrm L^2(\RR)$ with action \eqref{Bargmann_representation} \cite[§6g]{Bargmann:1954}. In either case it takes, of course, the Schrödinger equation to extract an irreducible subspace from the space of \emph{all} functions of $(\begin{smallmatrix}r\\t\end{smallmatrix})$.

(b): A state localized at $\{x_{|\LH_0}\}\subset\LH_0^*$ is one in which $q$ is certainly $\ell$. Again \eqref{nilpotentthm} provides the unique such state: $m(g) = \E^{-\I a}\E^{-\I\ell b}\d^c\sub0\d^e\sub0$. This turns out to be interesting. Indeed, computing as in \eqref{Heisenberg_modules} exhibits the resulting GNS module $\I(x,H_0)$ as $\ell^2(\RR^2)$ in which $G$ acts by the very formula \eqref{on_solutions}. By (\ref{polarization_dependance}a) this is irreducible even though $\mathrm I(x,H_0)$ is not. The need for Schrödinger's equation has evaporated!

The statistical interpretation sheds some light on this: inserting $m$ into (\ref{mu}), (\ref{nu_t}), we find
\begin{align}
   \textstyle
   \int_{\b\RR^2}
   \E^{\I\{(p,\gamma)-(E,\varepsilon)\}}
   \,d\mu(p,E)
   &=
   \d^\gamma_0\d^{\,\varepsilon\phantom{\gamma}}_0,
   &
   \textstyle
   \int_{\b\RR}
   \E^{-\I(r,\hspace{1pt}\beta)}
   \,d\nu_t(r)
   &=
   \left\{
   \begin{smallmatrix}
      \rlap{\small$\E^{-\I\ell\beta}$\quad if $t=0$}\\[3pt]
      \rlap{\small$\d^{\,\beta}_0$\hspace{3.5ex} else,}
   \end{smallmatrix}
   \right.
   \hspace{12ex}
\end{align}
i.e.~while $\nu_0$ is Dirac measure at $\ell$, both $\mu$ and $\nu_t$ $(t\ne0)$ are Haar measure. Thus we see that Theorem \eqref{nilpotentthm} gives Heisenberg's principle the form: \emph{position $q$ at any instant may be certain\textup, but then momentum-energy $(p,E)$ is necessarily equidistributed in the whole plane\textup, irrespective of the relation $E=\smash{\frac12p^2}$ in \eqref{paraboloid}}; and position $q+pt$ at any other instant is also equidistributed. 

This blurring of the relation $E=\smash{\frac12p^2}$ ``explains'', at the symbol level, the disappearance of Schrödin\-ger's equation. It is only under consideration here because, first, we do allow spectral measures concentrated on $\b X\suba$ and not just $X\suba$ \eqref{geomcrit2}, and secondly, the paraboloid \eqref{paraboloid} is Bohr dense in its affine hull (\ref{Howe-Ziegler}b). This may legitimate, in our opinion, the use of Bohr closure implicit in Definition \eqref{QR2}.
\end{myexam}

\section{Compact groups}

In this section $G$ is a compact connected Lie group. We fix a maximal torus $T\subset G$, and we write $W$ for the resulting Weyl group, $W=\textrm{Normalizer}(T)/T$. It is finite and acts on $\LT$ and $\LT^*$ by conjugation. We also fix a $W$-invariant inner product on $\LT$ and use it to identify $\LT$ and $\LT^*$. We have a canonical inclusion $\LT^*\hookrightarrow\LG^*$ as follows: being maximal abelian, $\LT$ coincides with the space of all $T$-fixed points in $\LG$; whence a canonical projection, $\int_T\operatorname{Ad}(t)\,dt:\LG\to\LT$, whose transpose identifies $\LT^*$ with the $T$-fixed points in $\LG^*$. We let:
\begin{enumerate}
   \item[(\hypertarget{root_system}{6.1})] 
   $R$ consist of the nonzero weights of $\LG_\CC$ (adjoint action), a.k.a.~roots;
   \item[(\hypertarget{Weyl_chamber}{6.2})] 
   $\overline C$ be the closure of a chosen connected component of $\LT\smallsetminus\bigcup_{\alpha\in R}\ker(\alpha)$;
   \item[(\hypertarget{order}{6.3})] 
   $\leqslant$ be defined on $\overline C$ by: $\lambda\leqslant\mu\Leftrightarrow\lambda$ is in the convex hull of $W(\mu)$ \cite[p.\,250]{Brocker:1985}.
\end{enumerate}
One knows:
\begin{itemize}
   \item[(\hypertarget{fundamental_domain}{6.4})]
   $\overline C$ is a fundamental domain for the $W$-action on $\LT=\LT^*$ \cite[p.\,202]{Brocker:1985};
   \item[(\hypertarget{X_inter_t*}{6.5})]
   each coadjoint orbit intersects $\LT^*$ in a $W$-orbit, hence $\overline C$ in a point \cite[p.\,74]{Bott:1979};
   \item[(\hypertarget{highest_weight}{6.6})]
   each irreducible continuous representation of $G$ has a $\leqslant$-highest weight in $\overline C$ 
   \linebreak\vspace{-\baselineskip}\item[\phantom{(6.6)}]
   which characterizes it \cite[p.\,252]{Brocker:1985}.
\end{itemize}

\addtocounter{equation}{6}
\begin{mytheo}
   \label{rep_lower_than_orbit}
   Every quantum representation of $G$ is continuous. The irreducible representation with highest weight $\lambda\in\overline C$ is quantum for the coadjoint orbit through $\mu\in\overline C$ if and only if $\lambda\leqslant\mu$ \textup{(\hyperlink{order}{6.3})}.
\end{mytheo}

\begin{proof}
   A unitary representation is continuous if and only if the state $m(g) = (\varphi,g\varphi)$ is continuous for each unit vector in it \cite[{}22.20a]{Hewitt:1963}. So it is enough to show that every quantum state (for $X$ say) is continuous. Now since $X$ is compact we have $\b X=X$, so for each abelian $\LA\subset\LG$ \eqref{geomcrit2} says that $m\comp\exp\suba$ has its spectral measure concentrated on $X\suba$ (in $\hat\LA$). By \cite[Korollar p.\,421]{Bauer:1959} it is equivalent to say that it is the image under $\LA^*\hookrightarrow\hat\LA$ of a measure $\nu$ concentrated on $X\suba$ (in $\LA^*$). So we have $(m\comp\exp\suba)(Z) = \textstyle\int_{\LA^*}\E^{\I\<u,Z\>}d\nu(u)$, which shows that $m\comp\exp\suba$ is continuous \eqref{Bochner}. Continuity of $m$ at $g\in G$ now follows by writing $\LG$ as a direct sum of lines $\LA_1,\dots,\LA_n$ and using the chart $(Z_1,\dots,Z_n)\mapsto g\exp(Z_1)\cdots\exp(Z_n)$, together with the inequality
   \begin{equation}
      \bigl|m(gg_1\cdots g_n) - m(g)\bigr|\leqslant \sqrt{2\Re(1 - m(g_1))}+\cdots+\sqrt{2\Re(1 - m(g_n))}
   \end{equation}
   which is obtained from \eqref{Krein} by induction on $n$.
   
   Suppose $\lambda\nleqslant\mu$. Let $V$ be the module with highest weight $\lambda$, and $X$ the orbit of $\mu$. If $\varphi\in V$ is a highest weight vector and  $m(g)=(\varphi,g\varphi)$, then $(m\comp\exp_{|\LT})(Z)=\E^{\I\<\lambda,Z\>}$ has its spectral measure concentrated at $\lambda\notin\Conv(W(\mu))$. But this convex hull is precisely $X_{|\LT}$ by Kostant's theorem (see e.g.~\cite{Ziegler:1992}), so Theorem \eqref{geomcrit2} says that $m$ and hence $V$ are not quantum for $X$.
      
   Conversely, suppose $\lambda\leqslant\mu$. Pick a unit vector $\varphi\in V$, write $m(g)=(\varphi,g\varphi)$ and let $E_\nu$ be the eigenprojector onto the subspace of weight $\nu$ vectors in $V$. Then $Z\in\LT$ acts on $V$ by $\sum_{\nu\,:\,\textrm{weight}}\I\<\nu,Z\>E_\nu$, so we have $(m\comp\exp_{|\LT})(Z) = \sum_{\nu\,:\,\textrm{weight}}\E^{\I\<\nu,Z\>}\|E_\nu\varphi\|^2$. Thus~the spectral measure of $m\comp\exp_{|\LT}$ is concentrated on the set of weights of $V$. Since these all lie in $\Conv(W(\lambda))\subset \Conv(W(\mu)) = X_{|\LT}$ by definition of $\leqslant$, we see that $m$ satisfies the condition of Theorem \eqref{geomcrit2} for $\LA=\LT$, for every unit $\varphi\in V$.
   
   But every maximal abelian subalgebra of $\LG$ is a \emph{conjugate} $\LA = g\inv\LT g$ of this one (e.g.~\cite[pp.\,73--74]{Bott:1979}). In that case, the obvious relation
   \begin{equation}
      (\varphi,\exp\suba(\,\cdot\,)\varphi) =
      (g\varphi,\exp_{|\LT}(\,\cdot\,)g\varphi)\comp\Ad(g)\suba
   \end{equation}
   shows that the spectral measure of 
   $(\varphi,\exp\suba(\,\cdot\,)\varphi)$
   is, dually, the image of the spectral measure of 
   $(g\varphi,\exp_{|\LT}(\,\cdot\,)g\varphi)$
   by the map $j:\LT^*\to\LA^*$ transpose to $\Ad(g)\suba:\LA\to\LT$. Since the latter measure is concentrated on $X_{|\LT}$ for every $g\varphi$ (by the previous case), it follows that the former is concentrated on $j(X_{|\LT})=X\suba$ for every $\varphi$, and we conclude by Theorem \eqref{geomcrit2} that $V$ is quantum for $X$.
\end{proof}
 
Theorem \eqref{rep_lower_than_orbit} shows that even in the compact case Definition \eqref{QR2} fails to recover the whole substance of the orbit method, which is (usually) understood to impose $\lambda = \mu$, i.e.~attach each representation to the orbit through its highest weight. While \cite{Ziegler:1996b} discusses various reasonable conditions one can add to regain this condition (e.g.~it suffices to restrict attention to modules weakly contained in sections of the Kostant-Souriau line bundle over the orbit \cite[Thm 5.23]{Ziegler:1996b}), we concentrate here on studying the representations obtained from states localized at an orbit $Y$ of a subgroup.

Although we mentioned after \eqref{localized} that the preimage of $Y$ in $X$ is \emph{generically} coisotropic, the useful case to consider below lies at the opposite end, where this preimage is a single point---as happens when we take $Y$ to be an extreme point (such as $X\cap\overline C$) of the convex polytope $X_{|\LT}$:

\begin{mytheo}
   \label{compactthm}
   Let $X$ be the coadjoint orbit through $\lambda\in\overline C$. If $\lambda$ is integral\textup, then there is a unique quantum state for $X$ localized at $\{\lambda_{|\LT}\}\subset\LT^*$\textup, namely $m(g)=(\varphi,g\varphi)$ where $\varphi$ is a highest weight vector in the irreducible $G$-module with highest weight $\lambda$. Otherwise there is no such state.
\end{mytheo}

\begin{proof}
   \smartqed
   Let $m$ be such a state, and write $\GNS{m}=\bigoplus_jV_{\lambda_j}$ for the (orthogonal) decomposition of the resulting GNS module \eqref{GNSS} into irreducibles with highest weights $\lambda_j$. Since $\GNS{m}$ is quantum for $X$ \eqref{GNS_quantum}, all $\lambda_j$ are $\leqslant\lambda$ \eqref{rep_lower_than_orbit}. Moreover we know that its cyclic vector $m_e$ (\hyperlink{cyclic_vector}{A.7}) is a weight vector of weight $\lambda$ \eqref{eigenvector}. So $\lambda$ must be integral, and $m_e$ is orthogonal to all summands with highest weights $\lambda_j<\lambda$, which must therefore vanish since $m_e$ is cyclic. Also by the orthogonality of vectors with different weight, $m_e$ is orthogonal to all except the maximal weight space in each remaining summand. So its decomposition writes $m_e = \sum_jc_j\varphi_j$ where $\varphi_j$ is a unit highest weight vector in $V_{\lambda_j}\cong V_\lambda$. Now the equivalence and orthogonality of the summands implies $(\varphi_j,g\varphi_k)=\d_{jk}(\varphi,g\varphi)$ where $\varphi$ is as in the statement of the Theorem. So we have
   \begin{equation}
      m(g) = 
      (m_e,gm_e)= 
      \sum_{j,k}\bar c_jc_k(\varphi_j,g\varphi_k) = 
      (\varphi, g\varphi),
   \end{equation}
   as claimed. (Of course it follows \emph{a posteriori} that there was only one summand.)
\end{proof}

\begin{myrema}
   Conjugating by a Weyl group element, \eqref{compactthm} will give a unique quantum state localized at any other extreme point of the polytope $X_{|\LT}$.
\end{myrema}

\section{Euclid's group and localization on normal congruences}

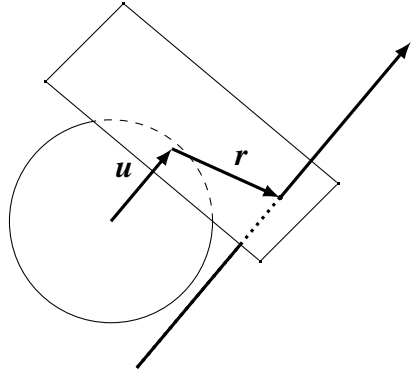
\begin{figure}[b!]
  \centering
  \begin{tikzpicture}[scale=0.9,inner sep=0pt, outer sep=0pt]
     \draw[dashed] [name path=dotted arc] (1.5cm,0) arc (0:100:1.5cm);
     \draw [name path=solid arc](1.5cm,0) arc (0:-260:1.5cm);
     \path [name intersections={of=dotted arc and solid arc}];
     \coordinate (A) at (intersection-1);
     \coordinate (B) at (intersection-2);
     \node [fill=black] (C) at ($ (A)!-.4!(B) $) {};
     \node [fill=black] (D) at ($ (A)!1.4!(B) $) {};
     \node (E) [fill=black,above right=of C] {};
     \node (F) [fill=black,above right=of D] {};
     \draw (C) -- (D) -- (F) -- (E) -- (C);
     \draw[-latex,very thick,outer sep=3pt] (0,0) to node[above left] {\large$\bm u$} (50:1.4);
     \coordinate (G) at (-80:2.2);
     \draw[very thick] (G) -- +(50:2.4);
     \draw[very thick,dotted] (G) -- +(50:3.3);
     \draw[fill] (2.5,0.35) circle (0.3mm);
     \draw[-latex,very thick] (2.5,0.35) -- +(50:3);
     \draw[-latex,very thick,outer sep=3pt] (50:1.4)  to node[above right] {\large$\bm r$} (2.5,0.35);
  \end{tikzpicture}
  \caption{Identification of the manifold $X$ of oriented lines (or light rays) with the tangent bundle $\mathrm{TS}^2$, after Hudson \cite{Hudson:1902}. Euclid's group acts on oriented lines via its natural action on $\RR^3$.}
  \label{fig:TS2}
\end{figure}

We consider here the manifold $X$ of oriented straight lines in Euclidean space $\RR^3$, i.e.~pairs $x=(\begin{smallmatrix}\ell\\\bm u\end{smallmatrix})$ of a line $\ell=\bm r +\RR\bm u$ and the choice $\bm u$ of one of the two unit vectors parallel to it. We can regard it either as the quotient of $\RR^3\times\mathrm S^2$ by the equivalence $(\begin{smallmatrix}\bm r\\\bm u\end{smallmatrix})\sim(\begin{smallmatrix}\bm r\smash{'}\\\bm u\smash{'}\end{smallmatrix})$ if $\bm u=\bm u'$ and $\bm r-\bm r'\parallel\bm u$, or as the subspace $\mathrm{TS}^2=\left\{(\begin{smallmatrix}\bm r\\\bm u\end{smallmatrix}): \bm r\perp\bm u\right\}$ which is a section of that quotient (Fig.~\ref{fig:TS2}). Either way, $X$ is naturally acted upon by Euclid's group $G$ (resp.~its Lie algebra $\LG$) consisting of all matrices of the form
\begin{equation}
   g=\begin{pmatrix}\,A\,&\,\bm c\,\\0&1\end{pmatrix},
   \qquad\text{resp.}\qquad
   Z=\begin{pmatrix}\,\j(\bm\alpha)\,&\bm\gamma\,\\0&0\,\end{pmatrix},
\end{equation}
where $A\in\mathrm{SO}(3)$, $\bm c,\bm\alpha,\bm\gamma\in\RR^3$ and $\j(\bm\alpha)=\bm{\alpha\times\cdot\,}$ (``vector product by $\bm\alpha$''). Moreover one can show that the most general $G$-invariant symplectic structure on $X$ writes 
\begin{equation}
   \omega(\d x,\d'x)=
   k\left[\<\d\bm u,\d'\bm r\> -\<\d'\bm u,\d\bm r\>\right]
   +s\<\bm u,\d'\bm u\bm\times\d\bm u\>
\end{equation}
for some $k>0$ and $s\in\RR$. (The term in $k$ was discovered by Lagrange \cite{Lagrange:1805} and the term in $s$ by Cartan \cite{Cartan:1896}.) Identifying $\LG^*$ with $\RR^6$ where $w=(\begin{smallmatrix}\mathbf L\\\mathbf P\end{smallmatrix})$ is paired to $Z\in\LG$ by $\<w,Z\> = \<\mathbf L,\bm\alpha\> + \<\mathbf P,\bm\gamma\>$, the resulting equivariant moment map $\Phi:X\to\LG^*$,
\begin{equation}
   \label{moment}
   \Phi(x)
   =\begin{pmatrix}\,\bm{r\times}k\bm u+ s\bm u\,\\ k\bm u\end{pmatrix},
\end{equation}
identifies $(X,\omega)$ with the coadjoint orbit $X^{k,s}$ of $(\begin{smallmatrix}s\bm e_3\\k\bm e_3\end{smallmatrix})$ endowed with its Kirillov-Kostant-Souriau 2-form. When so endowed, we think of $X$ as the manifold of \emph{\textbf{light rays with color $k$ and helicity $s$}}, and as the arena of geometrical optics \cite[{}15.88]{Souriau:1970}. In what follows we exhibit three kinds of lagrangian submanifolds (known classically as \emph{\textbf{normal congruences}}) on which light accepts to be concentrated:

\begin{figure}[h!]
  \centering
	\begin{tikzpicture}[scale=1.0]
	   \shadedraw[shading angle=270] (3.5,0) circle (4ex);
	   \node at (3.5,-1) {\eqref{convergent}: the zero};
	   \node at (3.5,-1.35) {section};
	   \node [draw,
	     shape=cylinder,
	     aspect=1.2,
	     minimum height=8ex,
	     minimum width=8ex,
	     left color=white!30,
	     right color=black!60,
	     middle color=gray!20,
	     outer sep=-0.5\pgflinewidth,
	     shape border rotate=90
	   ] at (7,0) {};
	   \node [draw,
	     shape=cylinder,
	     aspect=1.2,
	     minimum height=5.5ex,
	     minimum width=8ex,
	     left color=white!30,
	     right color=black!60,
	     middle color=gray!20,
	     outer sep=-0.5\pgflinewidth,
	     shape border rotate=90
	   ] at (7,-.35) {};
	   \draw[semithick] (7,0) circle (4ex);
	   \node at (7,-1) {\eqref{neon}: the equator's};
	   \node at (7,-1.35) {normal bundle};
	   \shadedraw (-1.5,.4) -- (1.5,.4) -- (1,.8) -- (-1,.8) -- (-1.5,.4);
	   \draw[semithick] (0,0) circle (4ex);
	   \filldraw (0,3.6ex) circle (.8pt);
	   \node at (0,-1) {\eqref{parallel}: the tangent space};
	   \node at (0,-1.35) {at the north pole};
	\end{tikzpicture}
\end{figure}

\begin{myexam}[Localization on a parallel beam]
   \label{parallel}
   Let $H$ be the subgroup of $G$ in which the rotation $A$ has axis $\RR\bm e_3$, i.e.
   $
   H=\bigl\{
   (\begin{smallmatrix}A & \bm c\\0&1\end{smallmatrix}):A=\E^{\,\j(\alpha\bm e_3)}\text{ for some }\alpha\in\RR
   \bigr\}
   $.
   Then $\bigl\{(\begin{smallmatrix}s\bm e_3\\k\bm e_3\end{smallmatrix})_{|\LH}\bigr\}$ is a point-orbit of $H$ in $\LH^*$, whose preimage in $X$ is the fiber $\mathrm T_{\bm e_3}\mathrm S^2\subset \mathrm{TS}^2$, i.e.~the lagrangian congruence of all lines normal to the plane $\bm e_3^\perp$. 
\end{myexam}

\begin{mytheo}
   \label{parallel_beam}
   If $s$ is an integer\textup, there is a unique quantum state for $X^{k,s}$ localized at $\bigl\{(\begin{smallmatrix}s\bm e_3\\k\bm e_3\end{smallmatrix})_{|\LH}\bigr\}\subset\LH^*$\textup, viz.
	\begin{equation}
	   \label{plane_wave}
	   m\begin{pmatrix}\,A\,&\,\bm c\,\\0&1\end{pmatrix}=
	   \begin{cases}
	      \ \ \E^{\I s\alpha}\E^{\I\<k\bm e_3,\bm c\>} & \text{if $A=\E^{\,\j(\alpha\bm e_3)}$\textup,}\\
	      \ \ 0 & \text{otherwise.}
	   \end{cases}
	\end{equation}
	The resulting $\GNS{}$ module \eqref{GNSS} is $\ind_H^G\chi^{k,s}$ where $\chi^{k,s}=m_{|H}$ and induction is in the sense of discrete groups\textup; it is irreducible. If $s$ is not an integer\textup, then there is no such state.
\end{mytheo}

\begin{proof}
    The fact that a localized state must coincide with \eqref{plane_wave} in $H$, and in particular that $s$ must be an integer, is just \eqref{point_orbits}. To see that it must vanish outside $H$, pick $g=(\begin{smallmatrix}A & *\\0&1\end{smallmatrix})\in G\smallsetminus H$ (thus $A\bm e_3\ne\bm e_3$) and then $h=(\begin{smallmatrix}\bm1 & \bm c\\0&1\end{smallmatrix})\in H$ such that $\E^{\I\<A\bm e_3 - \bm e_3,k\bm c\>}\ne1$. Computing as in \eqref{Weil_twice}, we get
    \begin{equation}
 	   \E^{\I\<\bm e_3,k\bm c\>}m(g)
 	   =m(hg)
 	   =m(gg\inv hg)
 	   =m(g)\E^{\I\<A\bm e_3,k\bm c\>}
    \end{equation}
    which shows that $m(g)=0$. The identification of $\GNS{m}$ as an induced representation is a special case of \eqref{Blattner}, and its irreducibility is a simple application of \eqref{Mackey-Shoda}. In fact, taking $\chi=\eta=m_{|H}$ there, the assignment $gH\mapsto A\bm e_3$ identifies $G/H$  with the sphere $\mathrm S^2$, on which the residual left action of $H$ is by rotations about $\RR\bm e_3$. So the only finite orbits (or double coset projections) are the poles $\pm\bm e_3$, and consequently the double cosets satisfying (\ref{Mackey-Shoda}b) are all contained in $H\+=\bigl\{(\begin{smallmatrix}A & \bm c\\0&1\end{smallmatrix}):A\bm e_3=\pm\bm e_3\bigr\}$ which is the normalizer of $H$. But if $g\in H\+$ projects to the south pole (so $A\bm e_3=-\bm e_3$) then we have just seen that $\chi(g\inv hg)$ could differ from $\chi(h)$. So the double cosets that also satisfy (\ref{Mackey-Shoda}a) are all contained in $\bigl\{(\begin{smallmatrix}A & \bm c\\0&1\end{smallmatrix}):A\bm e_3=\bm e_3\bigr\}$, which is just $H$. Hence the number of double cosets in \eqref{Mackey-Shoda} is just one, which shows that $\ind_H^G\chi^{k,s}$ is irreducible.
    
    There remains to show that the state \eqref{plane_wave} is indeed quantum for $X^{k,s}$. To this end we observe that $\LG$ has exactly two conjugacy classes of maximal abelian subalgebras. The first one consists of the translation ideal $\LT=\bigl\{(\begin{smallmatrix}0 & \bm\gamma\\0&0\end{smallmatrix}):\bm\gamma\in\RR^3\bigr\}$ alone. Identifying its dual with $\RR^3$ in the obvious way, it is clear on \eqref{moment} that $X^{k,s}{}_{|\LT}$ is the sphere of radius $k$, and on \eqref{plane_wave} that $m\comp\exp_{|\LT}$ has its spectral measure concentrated at its north pole $k\bm e_3$. So the condition of Theorem \eqref{geomcrit2} is satisfied. The other conjugacy class consists of the infinitesimal stabilizers
    \begin{equation}
       \label{stabilizer}
       \LG_x
       =
       \left\{
       Z=
       \alpha\begin{pmatrix}\,\j(\bm u)\, & \,\bm{r\times u}\,\\0&0\end{pmatrix}
       +
       \gamma\begin{pmatrix}\,0\,& \,\bm u\,\\0&0\end{pmatrix}
       :\alpha, \gamma\in\RR
       \right\}
    \end{equation}
    of all oriented lines $x=(\smash{\begin{smallmatrix}\bm r+\RR\bm u\\\bm u\end{smallmatrix}})\in X$. Identifying elements of $\LG_x^*$ with pairs $(\smash{\begin{smallmatrix}\ell\\p\end{smallmatrix}})$ so that $\<(\smash{\begin{smallmatrix}\ell\\p\end{smallmatrix}}),Z\>=\ell\alpha + p\gamma$ (so $\ell$ and $p$ are respectively the angular momentum \emph{around} and the linear momentum \emph{along} the oriented line $x$), one deduces readily from \eqref{moment} that the projection $X^{k,s}{}_{|\LG_x}$ is the strip $\left\{(\smash{\begin{smallmatrix}\ell\\p\end{smallmatrix}}): \ell\in\RR, -k<p<k\right\}$ with the two points $\pm(\smash{\begin{smallmatrix}s\\k\end{smallmatrix}})$ added. On the other hand \eqref{plane_wave} gives
    \begin{equation*}
       \label{m_stabilizer}
       (m\comp\exp_{|\LG_x})(Z)
       =
       m
       \begin{pmatrix}
          \E^{\,\j(\alpha\bm u)} & 
          (\mathbf 1-\E^{\,\j(\alpha\bm u)})\bm r+\gamma\bm u\\
          0 & 1
       \end{pmatrix}\\
       =
       \begin{cases}
          \ \E^{\pm\I (s\alpha+ k\gamma)} & \text{if $\bm u=\pm\bm e_3$}\\
          \ 1_{2\pi\ZZ}(\alpha)\E^{\I\gamma\<k\bm e_3,\bm u\>} & \text{otherwise,}\\
       \end{cases}
    \end{equation*}
    where $1_{2\pi\ZZ}$ is the characteristic function of $2\pi\ZZ$. In the first case we see that the spectral measure of $m\comp\exp_{|\LG_x}$ is Dirac measure at $\pm(\smash{\begin{smallmatrix}s\\k\end{smallmatrix}})$. In the second we see that it is Haar measure on $\b\ZZ\subset\b\RR$ \eqref{1_H} times Dirac measure at $\<k\bm e_3,\bm u\>$; so again the condition of Theorem \eqref{geomcrit2} is satisfied.
 \end{proof}

\begin{myremas}
   \label{simplification}
   (a) Although instructive, it is not actually necessary to check the condition of Theorem \eqref{geomcrit2} separately for $\LA=\LG_x$ as we have just done. Indeed, concentration of the spectral measure of $m\comp\exp_{|\LT}$ on the sphere $X^{k,s}{}_{|\LT}$ suffices to ensure concentration of the spectral measure of $m\comp\exp_{|\LG_x\cap\LT}$ on the segment $[-k,k]$ which is its image under  the projection $\smash{\hat\LT\to\widehat{\LG_x\cap\LT}}$; and by \cite[n$^{\textrm o}$ V.6.2, Cor.~2]{Bourbaki:1967b} this implies concentration of the spectral measure of $m\comp\exp_{|\LG_x}$ on the strip $\smash{\b X^{k,s}{}_{|\LG_x\cap\LT}}=\b\RR\times[-k,k]$ which is the preimage of $[-k,k]$ under the projection $\hat\LG_x\to\smash{\widehat{\LG_x\cap\LT}}$.
%
%

   (b) The module $\GNS{m}=\ind_H^G\chi^{k,s}$ and its cyclic vector have various realizations familiar in physics. It consists of $\ell^2$ sections of the $s$th tensor power of the tangent (complex line) bundle $\mathrm{TS}^2\to\mathrm S^2$, or in other words, functions $f:\mathrm{SO}(3)\to\CC$~satisfying $f(\E^{\,\j(\alpha\bm u_3)}U)=\E^{-\I s\alpha}f(U)$ and $\|f\|^2 = \sum_{\bm u_3\in\mathrm S^2}|f(U)|^2<\infty$, where $U=(\bm u_1 \bm u_2 \bm u_3)$; the group $G$ acts on them by
   \begin{equation}
   	(gf)(U)=\E^{\I\<\bm u_3,k\bm c\>}f(A\inv U).
   \end{equation}
   \emph{\textbf{Case $\bm{s=0}$.}} Here $f$ only depends on $U$ via $\bm u_3$. Putting $\psi(\bm r)=\sum_{\bm u_3\in\mathrm S^2}\E^{-\I\<\bm u_3,k\bm r\>}f(\bm u_3)$ one gets a Hilbert space of almost-periodic solutions of the Helmholtz equation $\varDelta\psi+k^2\psi=0$, with norm $\|\psi\|^2$ the Bohr mean of $|\psi|^2$, cyclic vector the ``plane wave'' $\psi(\bm r)=\E^{-\I kz}$ ($z=\<\bm e_3,\bm r\>$), and natural ``scalar field'' $G$-action:
   \begin{equation}
      \label{scalar_field}
      (g\psi)(\bm r)=\psi(A\inv(\bm r - \bm c)).
   \end{equation}
   \emph{\textbf{Case $\bm{s=1}$.}} Here $f$ has the form $f(U)=\<\bm u_1+\I\bm u_2,\bm b(\bm u_3)\>$ for a unique $\ell^2$ tangent vector field $\bm b$ on the sphere, on which $G$ acts by $(g\bm b)(\bm u)=\E^{\<\bm u,k\bm c\>J}A\bm b(A\inv\bm u)$ where $J$ is the sphere's standard complex structure, $J\d\bm u=\j(\bm u)\d\bm u$. Defining now $\mathbf F(\bm r)=(\mathbf{B}+\I\mathbf{E})(\bm r)=\sum_{\bm u\in\mathrm S^2}\E^{-\<\bm u,k\bm r\>J}(\bm b - \I J\bm b)(\bm u)$, one gets a Hilbert space of almost-periodic solutions of the reduced Maxwell equations \cites[{}(9) p.\,349]{Weber:1901}[{}(5.5)]{Bialynicki-Birula:2013}
   \begin{equation}
      \label{Maxwell}
      \left\{\ 
      \begin{aligned}
         \div \mathbf B = 0,\quad && \rot\mathbf B = k\mathbf B\rlap{,}\\
         \div \mathbf E = 0,\quad && \rot\mathbf E = k\mathbf E\rlap{,\footnotemark}
      \end{aligned}
      \right.
   \end{equation}
   \footnotetext{Helmholtz's equation $\varDelta\mathbf F+k^2\mathbf F=0$ follows, for on divergence-free vector fields the curl provides a square root (à la Dirac) of $-\varDelta=\rot\rot -\mathbf{grad}\div$.}%
   with cyclic vector the ``circularly polarized plane wave'' $\mathbf F(\bm r)=\E^{-\I kz}(\bm e_1-\I\bm e_2)$ and natural ``vector field'' $G$-action:
   \begin{equation}
      \label{vector_field}
	   (g\mathbf F)(\bm r)=A\mathbf F(A\inv(\bm r - \bm c)).
   \end{equation}
\end{myremas}

\begin{myexam}[Localization on a convergent beam]
   \label{convergent}
   Assume $s=0$ and let $K$ be the rotation subgroup of $G$, i.e.
   $
   K=\bigl\{
   (\begin{smallmatrix}A & 0\\0&1\end{smallmatrix}):A\in\mathrm{SO}(3)
   \bigr\}
   $.
   Then $\{0\}$ is a point-orbit of $K$ in $\LK^*$, 
   whose preimage in $X$ is the zero section $\mathrm S^2\subset \mathrm{TS}^2$, i.e.~the lagrangian congruence of all lines normal to a sphere centered at the origin.
\end{myexam}

\begin{mytheo}
   \label{convergent_beam}
   There is a unique quantum state for $X^{k,0}$ localized at $\{0\}\subset\LK^*$\textup, viz.
   \begin{equation}
      \label{spherical_wave}
  	   m\begin{pmatrix}\,A\,&\,\bm c\,\\0&1\end{pmatrix}=
  	   \frac{\sin\|k\bm c\|}{\|k\bm c\|}.
   \end{equation}
   The resulting $\GNS{}$ module \eqref{GNSS} is irreducible and is $\Ind_H^G\chi^{k,0}$\textup, where $H$ and $\chi^{k,0}$ are as in \eqref{parallel_beam}.
\end{mytheo}

\begin{proof}
   Localization at $\{0\}\subset\LK^*$ implies by \eqref{point_orbits} that $m_{|K}=1$. So Weil's formula \eqref{Weil} gives 
   $
   m\bigl(
   (\begin{smallmatrix}\bm1 & \bm c\\0&1\end{smallmatrix})
   (\begin{smallmatrix}A & 0\\0&1\end{smallmatrix})
   \bigr)
   =m\bigl(
   (\begin{smallmatrix}A & 0\\0&1\end{smallmatrix})
   (\begin{smallmatrix}\bm1 & \bm c\\0&1\end{smallmatrix})
   (\begin{smallmatrix}A\inv & 0\\0\phantom{\scriptscriptstyle{-1}}&1\end{smallmatrix})
   \bigr)
   =m
   (\begin{smallmatrix}\bm1 & \bm c\\0&1\end{smallmatrix})
   $, i.e.
   \begin{equation}
      \label{A-independence}
      m\begin{pmatrix}\,A\,&\,\bm c\,\\0&1\end{pmatrix}
      =
      m\begin{pmatrix}\,\bm1\,&\,A\bm c\,\\0&1\end{pmatrix}
      =
      m\begin{pmatrix}\,\bm1\,&\,\bm c\,\\0&1\end{pmatrix}.
   \end{equation}
   If further $m$  is quantum for $X^{k,0}$ and $\LT=\bigl\{(\begin{smallmatrix}0 & \bm\gamma\\0&0\end{smallmatrix}):\bm\gamma\in\RR^3\bigr\}$, then the compactness of the 2-sphere $X^{k,0}{}_{|\LT}$ implies as in the proof of \eqref{rep_lower_than_orbit} that
   $
   m(\begin{smallmatrix}\bm1 & \bm c\\0&1\end{smallmatrix})=
   \smash{\int_{\mathrm S^2}\E^{\I\<\bm u,k\bm c\>}d\nu(\bm u)}
   $
   for a unique probability measure $\nu$ on $\mathrm S^2$. Now the second equality in \eqref{A-independence} shows that $\nu$ has the rotation invariance property $\smash{\int_{\mathrm S^2} f(A\inv\bm u)\,d\nu(\bm u)}= \smash{\int_{\mathrm S^2} f(\bm u)\,d\nu(\bm u)}$ for all $f=\E^{\I\<\,\cdot\,,\,k\bm c\>}$. Since these span a uniformly dense subspace of the continuous functions on~$\mathrm S^2$ (Stone-Weierstrass) it follows that $\nu$ is the unique invariant probability measure on $\mathrm S^2$. Therefore we obtain, using spherical coordinates with pole at $\bm c/\|\bm c\|$,
   \begin{equation}
      \label{Poisson}
      m\begin{pmatrix}\,\bm1\,&\,\bm c\,\\0&1\end{pmatrix}=
      \frac1{4\pi}\int_0^{2\pi}\!\!\!\int_0^\pi\E^{\I\|k\bm c\|\cos\theta}\sin\theta\,d\theta\,d\varphi=
      \frac12\int_{-1}^1\E^{\I\|k\bm c\|z}dz=
      \frac{\sin\|k\bm c\|}{\|k\bm c\|}
   \end{equation}
   \cite[p.\,174]{Poisson:1820}. Together with \eqref{A-independence} this proves \eqref{spherical_wave}. Now consider the module $\Ind_H^G\chi^{k,0}\simeq\mathrm L^2(\mathrm S^2)$ with $G$-action $(gf)(\bm\varv)=\E^{\I\<\bm\varv,k\bm c\>}f(A\inv\bm\varv)$. It is irreducible by Mackey theory \cite[Thm 1]{Blattner:1965}, and we clearly have $m(g)=(f,gf)$ where $f(\bm\varv)\equiv1$. So \eqref{GNSS} shows that $m$ is a state and $\Ind_H^G\chi^{k,0}\simeq\GNS{m}$, as claimed. Finally it is clear from \eqref{Poisson} that $m\comp\exp_{|\LT}$ has its spectral measure concentrated on the sphere $\smash{X^{k,0}{}_{|\LT}}$, and from (\ref{simplification}a) that $m\circ\exp_{|\LG_x}$ has its own concentrated on the strip $\smash{\b X^{k,0}{}_{|\LG_x}}$. So we conclude by Theorem \eqref{geomcrit2} that $m$ is quantum for $\smash{X^{k,0}}$.
\end{proof}

\begin{myremas}
   %
   (a) For any integer $s$ one readily proves in the same manner that $\Ind_H^G\chi^{k,s}$ is irreducible and quantum for the orbit $X^{k,s}$. But only in the case $s=0$ do we have a characterization of this representation as arising from a localized state.
   
   \label{wave_equations2}
   (b) Just as the $\ind_H^G\chi^{k,s}$ can be realized in solution spaces of wave equations on $\RR^3$ (\ref{scalar_field}--\ref{vector_field}), so can the $\Ind_H^G\chi^{k,s}$: simply replace $\sum_{\bm u_3\in\mathrm S^2}$ there by $\int_{\mathrm S^2}\dots d\nu(\bm u_3)$. (The resulting norms on solution spaces are computed in \cite[Thm 5.5]{Strichartz:1990}.) In particular the cyclic vector $f(\bm\varv)\equiv1$ of $\Ind_H^G\chi^{k,0}$ becomes the ``spherical wave'' $\psi(\bm r)=\frac{\sin\|k\bm r\|}{\|k\bm r\|}$.
\end{myremas}

\begin{myexam}[Localization on a neon beam]
   \label{neon}
   Let 
   $
   G_a=\exp\bigl\{
   (\begin{smallmatrix}\j(\alpha\bm e_3) & \gamma\bm e_3\\0&1\end{smallmatrix}):\alpha, \gamma\in\RR
   \bigr\}
   $ be the stabilizer of the vertical axis $a=(\smash{\begin{smallmatrix}\RR\bm e_3\\\bm e_3\end{smallmatrix}})\in X$.
   Then $\{0\}$ is a point-orbit of $G_a$ in $\LG_a^*$, 
   whose preimage in $X\simeq\mathrm{TS}^2$ is the normal bundle to the equator $\mathrm S^1\subset \mathrm{S}^2$, i.e.~the lagrangian congruence of all lines normal to a cylinder with directrix $a$.
\end{myexam}

\begin{mytheo}
   \label{neon_beam}
   There are \textup(at least\textup) two pure quantum states for $X^{k,0}$ localized at $\{0\}\subset\LG_a^*$\textup, viz.
   \begin{equation}
      \label{cylindrical_wave}
  	   m_\varepsilon\begin{pmatrix}\,A\,&\,\bm c\,\\0&1\end{pmatrix}=
	   \begin{cases}
         \ \ J_0\left(\|k\bm c_\perp\|\right)
	      & \text{if $A\bm e_3=\bm e_3$\textup,}\\
         \ \ (-1)^\varepsilon J_0\left(\|k\bm c_\perp\|\right)
	      & \text{if $A\bm e_3=-\bm e_3$\textup,}\\
	      \ \ 0 & \text{otherwise\textup,}
	   \end{cases}
	   \rlap{\qquad$(\varepsilon=0,1)$\textup,}
   \end{equation}
   where $J_0$ is the zeroth-order Bessel function and $\bm c_\perp\!=$  projection of $\bm c$ in the plane~$\smash{\bm e_3^\perp}$. We have $\GNS{m_\varepsilon}=\ind_{H\+}^G\Ind_{T\+}^{H\+}\chi_\varepsilon$ where $\chi_\varepsilon(\begin{smallmatrix}A & \bm c\\0&1\end{smallmatrix})=(\pm 1)^\varepsilon\E^{\I\<k\bm e_1,\bm c\>}$ if $A\bm e_3=\pm\bm e_3$ and
   \begin{equation}
      H\+=\bigl\{(\begin{smallmatrix} A & \bm c\\0&1\end{smallmatrix})\in G:A\bm e_3=\pm\bm e_3\bigr\},
      \quad
      T\+=\bigl\{(\begin{smallmatrix} A & \bm c\\0&1\end{smallmatrix})\in G:A\in\{\bm 1, \E^{\,\j(\pi\bm e_1)}\}\bigr\}.
   \end{equation}
\end{mytheo}

\begin{proof}
   Let $m$ be a quantum state for $X^{k,0}$. As in the proof of \eqref{convergent_beam}, we have a probability measure $\lambda$ on $\mathrm S^2$ such that $\smash{m(\begin{smallmatrix} \bm1 & \bm c\\0&1\end{smallmatrix})=\int_{\mathrm S^2}\E^{\I\<k\bm u,\bm c\>}d\lambda(\bm u)}$. Localization at $\{0\}\subset\LG_a^*$ further implies that $m$ is trivial on $G_a$ and in particular on $\exp(\begin{smallmatrix} 0 & \RR\bm e_3\!\\0&0\end{smallmatrix})$. Writing $\pi$ for the projection $\bm u\mapsto ku_3$, it follows that the image $\pi(\lambda)$ is Dirac measure at $0$, hence that $\lambda$ is concentrated on the equator $\mathrm S^1\subset\mathrm S^2$ \cite[n$^{\textrm o}$ V.6.2, Cor.~4]{Bourbaki:1967b}. Next, the triviality of $m(\begin{smallmatrix} A & 0\\0&1\end{smallmatrix})$, $A\in\mathrm{SO}(2):=\{\E^{\,\j(\alpha\bm e_3)}:\alpha\in\RR\}$, implies that the relations \eqref{A-independence} hold for $A\in\mathrm{SO}(2)$ with the same proof. Therefore $\lambda$ is the $\mathrm{SO}(2)$-invariant measure on $\mathrm S^1$ and we have, with $H=\{(\begin{smallmatrix} A & \bm c\\0&1\end{smallmatrix})\in G:A\in\mathrm{SO}(2)\}$ as before,
   \begin{equation}
      \label{Parseval}
      m_{|H}\begin{pmatrix}\,A\,&\,\bm c\,\\0&1\end{pmatrix}=
      m_{|H}\begin{pmatrix}\,\bm1\,&\,\bm c\,\\0&1\end{pmatrix}=
      \int_{\mathrm S^1}\E^{\I\<\bm u,k\bm c_\perp\>}d\lambda(\bm u)=
      J_0(\|k\bm c_\perp\|)
   \end{equation}
   \cite[§2.2]{Watson:1922}. This shows that the restriction $m_{|H}$ must be given by the first row of \eqref{cylindrical_wave}.
      
   We \emph{do not know} whether the next two rows give the only extensions of the first row to pure states of $G$; but we can prove that they do provide such extensions. Indeed, consider the module $V_\varepsilon=\Ind_{T\+}^{H\+}\chi_\varepsilon\simeq\mathrm L^2(\mathrm S^1)$ with $H\+$\nobreakdash-action $(gf)(\bm u)=(\pm1)^\varepsilon\E^{\I\<\bm u,k\bm c\>}f(A\inv\bm u)$ whenever $A\bm e_3=\pm\bm e_3$. It is irreducible by Mackey theory \cite[Thm 1]{Blattner:1965} and we clearly have $m_\varepsilon{}_{|H\+}(g)=(f,gf)$ where $f(\bm u)\equiv1$. So \eqref{GNSS} shows that $m_{\varepsilon|H\+}$ is a state and $V_\varepsilon=\smash{\GNS{m_{\varepsilon|H\+}}}$. Now \cite[Thm~1]{Blattner:1963} says that the extension $m_\varepsilon$ of $m_{\varepsilon|H\+}$ by zero \eqref{cylindrical_wave} is a state and $\GNS{m_\varepsilon}=\ind_{H\+}^GV_\varepsilon$. Moreover we can show that the latter induced representation is irreducible. In fact \cite[Cor.~1]{Blattner:1962} proves that
   \begin{equation}
      \label{intertwining_number}
      \dim(\Hom_G(\ind_{H\+}^G\!V_\varepsilon,\ind_{H\+}^G\!V_\varepsilon))
      \leqslant\!\!\!\!
      \sum_{H\+gH\+\in H\+\backslash G/H\+}\!\!\!\!
      \dim(\Hom_{H\+\cap gH\+g\inv}(V_\varepsilon,{}^gV_\varepsilon)),
   \end{equation}
   where ${}^gV_\varepsilon$ denotes the $gH\+g\inv$-module in which $k\in gH\+g\inv$ acts as $g\inv kg$ acts~on~$V_\varepsilon$. Now if $g\in H\+$, then its double coset $H\+gH\+=H\+$ clearly contributes 1 to the sum in \eqref{intertwining_number}. On the other hand if $g\notin H\+$, then $H\+\cap gH\+g\inv$ contains the translation group $T$. But any $I\in\Hom_T(V_\varepsilon,{}^gV_\varepsilon)$ satisfies by definition $I\E^{\I\<\cdot,kc\bm e_3\>}f=\E^{\I\<\cdot,kcA\bm e_3\>}If$, or in other words (since the left-hand side here is just $If$)
   \begin{equation}
      (1-\E^{\I\<\bm u,kcA\bm e_3\>})(If)(\bm u)=0
      \qquad\forall\,c\in\RR.
   \end{equation}
   As $A\bm e_3\ne\pm\bm e_3$, the first factor is only zero (for all $c$) at two points of the equator, and we conclude that $I=0$. So the sum in \eqref{intertwining_number} is 1 and $\ind_{H\+}^GV_\varepsilon$ is irreducible; hence $m_\varepsilon$ is pure, as claimed. Finally it is clear from \eqref{Parseval} that $m_\varepsilon\comp\exp_{|\LT}$ has its spectral measure concentrated on (the equator of) the sphere $\smash{X^{k,0}{}_{|\LT}}$, and from (\ref{simplification}a) that $m_\varepsilon\circ\exp_{|\LG_x}$ has its own concentrated on the strip $\smash{\b X^{k,0}{}_{|\LG_x}}$. So we conclude by Theorem \eqref{geomcrit2} that $m_\varepsilon$ is quantum for $\smash{X^{k,0}}$.
\end{proof}

\begin{myremas} (a) As emphasized during the proof, we do not know if \eqref{cylindrical_wave} gives the \emph{only} pure quantum states for $X^{k,0}$ (or $X^{k,s}$) localized at $\{0\}\subset\LG_a^*$.
   
   (b) Much as in (\ref{simplification}b) and (\ref{wave_equations2}b), one can realize the representation $\GNS{m_0}$ in a Hilbert space of solutions of  $\varDelta\psi+k^2\psi=0$, with cyclic vector the ``cylindrical wave'' $\psi(\bm r)=J_0(\|k\bm r_\perp\|)$ and norm $\|\psi\|^2= \lim_{R\to\infty}R^{-2}\int_{\|\bm r\|\leqslant R}|\psi(\bm r)|^2d^3\bm r$ \cite[Thm 5.5]{Strichartz:1990}. On the other hand, we have not managed to produce a similar realization of $\GNS{m_1}$.
   
   (c) The modules $\ind_H^G\chi^{k,0}$ \eqref{parallel_beam} and $\Ind_H^G\chi^{k,0}$ \eqref{convergent_beam} were given by 
   the $G$-action
    $(gf)(\bm\varv)=\E^{\I\<\bm\varv,k\bm c\>}f(A\inv\bm\varv)$
   in $\mathrm L^2(\mu_0)$ and $\mathrm L^2(\mu_2)$, where $\mu_d$ is $d$-dimensional Hausdorff measure on the sphere. It would be interesting to determine if the same action in $\mathrm L^2(\mu_d)$ is also irreducible, and in particular if $\mathrm L^2(\mu_1)$ is isomorphic to  $\GNS{m_0}$ \eqref{neon_beam}.   
\end{myremas}

\begin{appendices}
   \renewcommand\thesection{\Alph{section}}
   
   \section{Appendix: Positive-definite functions, states, representations}
   
   \begin{mydefis}
      \label{state}
   	Let $G$ be a group, with identity element $e$. Recall that a complex-valued function $m$ on $G$ is called \emph{\textbf{positive-definite}} if the sesquilinear form
	   \begin{equation}
	      \label{sesquilinear}
	   	(c,d)_m := \sum_{g,h\in G}\overline c_g d_h m(g\inv h),
	   \end{equation}
	   defined on $\CC[G] = \{\text{complex-valued functions with finite support on $G$}\}$, is positive: $(c,c)_m\geqslant 0$. If further $m(e) = 1$, then $m$  is called a \emph{\textbf{state of $G$}}. A state of $G$ is called \emph{\textbf{pure}} (or \emph{\textbf{extreme}}) if it is not a convex combination of two states other than itself.
   \end{mydefis}

   We can identify each function $m$ on $G$ with the linear functional on $\CC[G]$ defined by $m(\d^g) = m(g)$, where $\d^g$ denotes the basis function which is one at $g$ and zero elsewhere; then \eqref{sesquilinear} writes $(c,d)_m = m(c^*\!\cdot d)$, where we are using the ${}^*$-algebra structure of $\CC[G]$: $\d^g\!\cdot\d^h =\d^{gh}$, $\d^{g*}=\smash{\d^{g\inv}}$. So states are the same as normalized positive linear functionals on $\CC[G]$.
   
   \begin{mytheo}[Gel'fand-Na{\u\i}mark-Segal, Schwartz \cite{Schwartz:1964}]
      \label{GNSS}
      A function $m$ on $G$ is a state if and only if there are a unitary $G$-module $\mathcal{H}$\textup, and a unit vector $\varphi\in\mathcal{H}$\textup, such that 
      \begin{equation}
         \label{sandwich}
      	m(g)=(\varphi,g\varphi).
      \end{equation}
       We may even assume that $\varphi$ is cyclic\textup, i.e.~its $G$-orbit spans a dense subspace of $\mathcal{H}$. Then the pair $(\mathcal{H},\varphi)$ is unique and canonically isomorphic to $(\GNS{m}, m_e)$\textup, where
       \begin{itemize}
          \item[\textup{(\hypertarget{GNS_m}{A.5})}]
          $\GNS{m}\subset\CC^G$ is the subspace with reproducing kernel $K(g,h)=m(g\inv h)$\textup;
          \item[\textup{(\hypertarget{G-action}{A.6})}]
          $G$ acts on it by $(gf)(g')=f(g\inv g')$\textup;
          \item[\textup{(\hypertarget{cyclic_vector}{A.7})}]
          the cyclic vector $m_e$ is the complex conjugate $\overline m=K(\,\cdot\,,e)$ of $m$\textup.
       \end{itemize}
       Finally $m$ is pure if and only if\/ $\GNS{m}$ is irreducible.
   \end{mytheo}
   \addtocounter{equation}{3}   

   \begin{proof}
      If \eqref{sandwich} holds, we get $m(e)=1$ and $m(c^*\!\cdot c)=(c\varphi,c\varphi)\geqslant 0$; so $m$ is a state. Conversely if $m$ is a state, one observes that the form \eqref{sesquilinear} on $\CC[G]$ is invariant under the regular action, $gc=\d^g\!\cdot c$; dividing out the null vectors $\CC[G]^\perp$ and completing, one obtains a unitary $G$-module $\overline{\CC[G]/\CC[G]^\perp}$ in which \eqref{sandwich} holds with $\varphi$ the class of $\d^e$.

      The clever way to complete here is to take the \emph{antidual} \cite{Schwartz:1964}: we let $\GNS{m}$ be the (contragredient) $G$-module consisting of all antilinear functionals $f$ on $\CC[G]$, such that the quantity
	   \begin{equation}
	      \label{GNS_norm}
	      \|f\|^2:=
	      \smash[t]{\sup_{c\in\CC[G]}\frac{|f(c)|^2}{(c,c)_m}}
	      \qquad
	      \textup{is finite.}
	   \end{equation}
      (It is understood that the numerator must vanish when the denominator does, so that $f$ factors through the null vectors.) Clearly each $d\in\CC[G]$ defines an element $m_d := (\,\cdot\,,d)_m$ of $\GNS{m}$, and one verifies without trouble that $d\mapsto m_d$ induces a $G$-equivariant linear isometry of $\CC[G]/\CC[G]^\perp$ into $\GNS{m}$; whence by extension an isometry $\overline{\CC[G]/\CC[G]^\perp}\to\GNS{m}$ which is onto by the Riesz representation theorem. In particular we have $(c,d)_m=(m_c,m_d)$ and thus (first for $f=m_d$, then in general by density) the ``reproducing'' property
	   \begin{equation}
	      \label{repro_kernel}
	      f(c)=(m_c,f)
	      \qquad
	      \forall\,f\in\GNS{m}
	   \end{equation}
	   of the kernel $K(\,\cdot\,,c):=m_c(\,\cdot\,)$. Now abbreviate $f(\d^g)$ to $f(g)$ and $m_{\d^g}$ to $m_g$: in~this way $\GNS{m}$ becomes a unitary $G$-module of functions on $G$, with reproducing kernel $K(g,h)=m_h(g)=m(g\inv h)$ and cyclic vector $m_{\d^e}=m_e$. Finally if $\varphi$ in \eqref{sandwich} is cyclic, then the map $c\varphi\mapsto m_c$ extends to the required isomorphism $\mathcal{H}\to\GNS{m}$; and for the equivalence $m$ pure $\Leftrightarrow$ $\GNS{m}$ irreducible we refer to \cite[{}21.34]{Hewitt:1963}.
   \end{proof}
   
   Before further exemplifying this construction, we record an important inequality \eqref{Weil} of Weil \cite[p.\,57]{Weil:1940} and some of its consequences:  

   \begin{mytheo}
      \label{inequalities}
      Every state satisfies $m(g\inv)=\overline{m(g)}$ and
      \begin{gather}
         \label{Herglotz}
         |m(g)|\leqslant 1,\\
         \label{Krein}
         \textstyle
         |m(g) - m(h)|\leqslant \sqrt{2\Re(1 - m(g\inv h))},\\
         \label{Weil}
         \textstyle
         |m(gh) - m(g)m(h)|\leqslant \sqrt{1 - |m(g)|^2}\sqrt{1 - |m(h)|^2}.
      \end{gather}
   \end{mytheo}
   
   \begin{proof}
      The first statement is because $(\d^g,\d^e)_m=\overline{(\d^e,\d^g)}{}_m$ since \eqref{sesquilinear} is \mbox{hermitian}. As it is positive we also have a Cauchy-Schwarz inequality: $|(c,d)_m|^2\!\leqslant(c,c)_m(d,d)_m$. This becomes \eqref{Herglotz} if we take the pair $c^*$, $d$ to be $\d^e$, $\d^g$; \eqref{Krein} if we take it to be $\d^e$, $\d^g-\d^h$; and \eqref{Weil} if we take it to be $\d^g-m(g)\d^e$, $\d^h-m(h)\d^e$.
   \end{proof}
   
   \begin{mycoro}
      For any state $m$ of $G$\textup, the equation $|m(g)|=1$ defines a subgroup $H$ of $G$\textup, $m$ restricts to a character $\chi$ of $H$\textup, and we have
      \begin{equation}
         \label{H-equivariance}
         f(gh)=\overline{\chi}(h)f(g)
         \qquad
         \forall\,(f,g,h)\in\textup{GNS}_m\times G\times H.
      \end{equation}
   \end{mycoro}
   
\begin{proof}
    The initial statements are clear from \eqref{Weil}. For \eqref{H-equivariance}, let \mbox{$d=\d^h-m(h)\d^e$}. Then $\|m_{gd}\|^2=(d,d)_m=0$, whence $f(gh)-\overline{\chi}(h)f(g)=f(gd)=0$ by \eqref{repro_kernel}.
\end{proof}

   Property \eqref{H-equivariance} means that $\GNS{m}$ is a certain space of sections of the line bundle, $G\times_H\CC$, associated to $G\to G/H$ by the character $\chi$. Which space exactly, and with what norm, depend on how $m$ extends $\chi$ off $H$. For instance, we will show that we get all $\ell^2$ sections if we take the extension by \emph{zero}, i.e.~the state
   \begin{equation}
      \label{chi^bullet}
      m(g)=\chi^\bullet(g) = 
	   \begin{cases}
	      \ \ \chi(g) & \text{if $g\in H$,}\\
	      \ \ 0 & \text{otherwise.}
	   \end{cases}
   \end{equation}

	\begin{mytheo}[Blattner \cite{Blattner:1963}]
      \label{Blattner}
      For $m=\chi^\bullet$ as above\textup, we have $\GNS{m}=\ind_H^G\chi$ where induction is in the sense of discrete groups. That is to say\textup, the space \eqref{GNS_norm} consists exactly of all $f:G\to\CC$ such that
      \begin{itemize}
	      \item[\textup{(a)}] $f(gh)=\overline{\chi}(h)f(g)$ for all $h\in H;$
	      \item[\textup{(b)}] the quantity $\|f\|^2_\star:=\sum_{gH\in G/H}|f(g)|^2$ is finite.
      \end{itemize}
   \end{mytheo}
         
   \begin{proof}
      First we confirm that \eqref{chi^bullet} is positive-definite: splitting the sum \eqref{sesquilinear} over the cosets of $H$ one readily obtains $(c,c)_m=\sum_{gH\in G/H}\left|m_c(g)\right|^2\geqslant0$, where $m_c(g)=\sum_{h\in H}c_{gh}\chi(h)$ is the function defined before \eqref{repro_kernel}.
       
       Assume that $f$ satisfies \eqref{GNS_norm}. Then \eqref{H-equivariance} proves (a), and taking $c = \sum_{g\in \Gamma} f(g)\d^g$ where $\Gamma\subset G$ is finite with at most one point per $H$-coset, one finds that the quotient in \eqref{GNS_norm} equals $\sum_{g\in \Gamma}|f(g)|^2$. This shows that $\|f\|_\star^2\leqslant\|f\|^2$, whence (b).

       Conversely, assume that $f$ satisfies (a,\,b). Splitting the sum $f(c)=\sum_{g\in G}\overline c_gf(g)$ over the cosets of $H$ gives $f(c)=\sum_{gH\in G/H}\overline{m_c}(g)f(g)$. Inserting this and the above value of $(c,c)_m$ into \eqref{GNS_norm}, and using Cauchy-Schwarz, one obtains $\|f\|^2\leqslant\|f\|^2_\star$.
   \end{proof}
      
   The realization \eqref{GNS_norm} is especially well suited to discuss intertwining operators $J:\GNS{m}\to\GNS{n}$, for each will be characterized by a single function, $Jm_e$. In more detail, writing $\ch$ for the involution $f\mapsto f\ch:=\smash{\overline{f(\cdot\,{}^{\raisebox{-1pt}{\scriptsize$-1$}})}}$ of $\CC^G$, we have:

   \begin{mytheo}
      \label{J_to_j}
      Let $m$\textup, $n$ be two states of $G$. Then $J\mapsto Jm_e$ defines an injection
      $
      	\Hom_G(\GNS{m},\GNS{n})
      	\longrightarrow
      	\operatorname{GNS}_{\smash{\raisebox{1pt}{$\scriptstyle m$}}}\ch\cap\GNS{n}.
      $
   \end{mytheo}
   
   \begin{proof}
      By hypothesis the function $j=Jm_e$ is in $\GNS{n}$ and satisfies $gj=Jm_g$. Thus, by \eqref{repro_kernel}, the adjoint of $J$ is given by $(J^*f)(g)=(m_g,J^*f)=(gj,f)$. In particular, putting $f=n_e$ one finds $J^*n_e=j\ch$. Therefore $j\ch$ is in $\GNS{m}$, and it determines $J$ by the dual calculation: $(Jf)(g)=(n_g,Jf)=(J^*n_g,f)=(gj\ch,f)$.
   \end{proof}

   \begin{mycoro}[Mackey-Shoda {\cite[{}II.2]{Mackey:1951}}]
      \label{Mackey-Shoda}
      Let $\chi$ and $\eta$ be characters of subgroups $H$ and $K$ of $G$. Then $\Hom_G(\ind_H^G\chi,\ind_K^G\eta)$ has its dimension bounded above by the number of double cosets $D=HgK$ such that
      \begin{itemize}
	      \item[\textup{(a)}] $\chi(h)=\eta(g\inv hg)$ for all $\,h\in H\cap gKg\inv;$
	      \item[\textup{(b)}] $HgK$ projects onto finite sets in both $G/K$ and $H\backslash G$.
      \end{itemize}
   \end{mycoro}
   
   \begin{proof}
      By \eqref{J_to_j} this dimension does not exceed that of $(\ind_H^G\chi)\ch\cap(\ind_K^G\eta)$, whose members $j$ satisfy $j(h\inv gk)=\overline{\eta}(k)j(g)\chi(h)$ by virtue of (\ref{Blattner}a).

      Such a function is determined by one value per double coset $D=HgK$. This value must vanish when (a) fails, as one sees by putting $k=g\inv hg\,$ in the relation above; also when (b) fails: for $|j|^2$ is constant in $D$, and this constant occurs $\sharp(D/K)$ times in the series (\ref{Blattner}b) for $\|j\|^2$, resp.~$\sharp(H\backslash D)$ times in the series for $\|j\ch\|^2$.
   \end{proof}
   
   We conclude this Appendix with Bochner's description of continuous positive-definite functions on \emph{locally compact abelian} groups \cite[pp.\,120--122]{Weil:1940}. If $G$ is such a group, write $\hat G$ for its Pontryagin dual, i.e.~the group of all continuous characters $\chi:G\to\mathrm U(1)$ with the topology of uniform convergence on compact sets.
   \begin{mytheodefi}[Bochner]
      \label{Bochner}
      The Fourier transformation $\mu\mapsto m$\textup:
      \begin{equation}
         m(g) = \textstyle\int_{\hat G}\chi(g)\,d\mu(\chi)
      \end{equation}
      defines a bijection between all continuous positive-definite functions $m$ on $G$\textup, and all positive bounded Radon measures $\mu$ on $\hat G$. In particular\textup, states of $G$ correspond to probability measures on $\hat G$. We refer to $\mu$ as the \textbf{spectral measure} of $m$. 
   \end{mytheodefi}
   
   \begin{myexam}
      In the setting of \eqref{Bochner}, suppose that $H$ is an \emph{open subgroup}~of~$G$. The characteristic function $1_H$ of $H$ in $G$ is a continuous state of $G$ \eqref{chi^bullet}, and we claim that its spectral measure is the image of Haar measure on the annihilator $H^\perp=\bigl\{\chi\in\hat G:\chi(h)=1 \text{ for all } h\in H\bigr\}$ under the inclusion $H^\perp\hookrightarrow\hat G$, i.e.~we have
      \begin{equation}
         \label{1_H}
         1_H(g)=\textstyle\int_{H^\perp}\eta(g)\,d\eta.
      \end{equation}
      To prove this, we first observe that $H$ is also closed (as complement of the union of its cosets in $G$); so $G/H$ is discrete and its dual $\smash{\widehat{G/H}}\simeq H^\perp$ is compact \cite[{}23.17, 23.25, 23.29]{Hewitt:1963}. So Haar measure $d\eta$ on $H^\perp$ is a probability measure, and the right-hand side $m(g)$ of \eqref{1_H} is clearly 1 when $g\in H$. On the other hand, the translation invariance of $d\eta$ gives $m(g) = \int_{H^\perp}(\zeta\eta)(g)\,d\eta = \zeta(g)m(g)$ for all $\zeta\in H^\perp$. If $g\notin H$ this implies $m(g)=0$, for we can find $\zeta\in H^\perp$ such that $\zeta(g)\ne 1$ \cite[{}23.26]{Hewitt:1963}.
   \end{myexam}
   
\end{appendices}   
\setlength{\labelalphawidth}{2.8em}

\let\l\polishl

\nocite{Souriau:2003}

\currentpdfbookmark{References}{References}
\phantomsection{}
\mtaddtocont{\protect\contentsline{mtchap}{\protect\numberline{}References}{\thepage}\hyperhrefextend}
\printbibliography
\end{document}